\documentclass[aps, prl, reprint, superscriptaddress]{revtex4-1}

\usepackage{todonotes}
\usepackage{bm}
\usepackage{graphicx}
\usepackage{amsmath}
\usepackage{color}
\usepackage{pdfpages}

\begin{document}

\title{Ground state search, hysteretic behaviour, and reversal mechanism of skyrmionic textures in confined helimagnetic nanostructures}

\author{Marijan Beg}
\email{mb4e10@soton.ac.uk}
\author{Rebecca Carey}
\author{Weiwei Wang}
\author{David Cort\'{e}s-Ortu\~{n}o}
\author{Mark Vousden}
\author{Marc-Antonio Bisotti}
\author{Maximilian Albert}
\author{Dmitri Chernyshenko}
\author{Ondrej Hovorka}
\affiliation{Faculty of Engineering and the Environment, University of Southampton, Southampton SO17 1BJ, United Kingdom}
\author{Robert L. Stamps}
\affiliation{SUPA School of Physics and Astronomy, University of Glasgow, Glasgow G12 8QQ, United Kingdom}
\author{Hans Fangohr}
\email{h.fangohr@soton.ac.uk}
\affiliation{Faculty of Engineering and the Environment, University of Southampton, Southampton SO17 1BJ, United Kingdom}

\begin{abstract}
Magnetic skyrmions have the potential to provide solutions for low-power,
high-density data storage and processing. One of the major challenges in
developing skyrmion-based devices is the skyrmions' magnetic stability in
confined helimagnetic nanostructures. Through a systematic study of
equilibrium states, using a full three-dimensional micromagnetic model
including demagnetisation effects, we demonstrate that skyrmionic textures are
the lowest energy states in helimagnetic thin film nanostructures at zero
external magnetic field and in absence of magnetocrystalline anisotropy. We
also report the regions of metastability for non-ground state equilibrium
configurations. We show that bistable skyrmionic textures undergo hysteretic behaviour
between two energetically equivalent skyrmionic states with different core
orientation, even in absence of both magnetocrystalline and demagnetisation-based
shape anisotropies, suggesting the existence of Dzyaloshinskii-Moriya-based
shape anisotropy. Finally, we show that the skyrmionic texture core
reversal dynamics is facilitated by the Bloch point occurrence and
propagation.
\end{abstract}

\maketitle 
An ever increasing need for data storage creates great challenges
for the development of high-capacity storage devices that are cheap, fast,
reliable, and robust. Nowadays, hard disk drive technology uses magnetic
grains pointing up or down to encode binary data (0 or 1) in so-called
perpendicular recording media. Practical limitations are well understood and
dubbed the ``magnetic recording trilemma''~\cite{Richter2007}. It defines a
trade-off between three conflicting requirements: signal-to-noise ratio,
thermal stability of the stored data, and the ability to imprint information.
Because of these fundamental constraints, further progress requires radically
different approaches.

Recent research demonstrated that topologically stable magnetic skyrmions have
the potential for the development of future data storage and information
processing devices. For instance, a skyrmion lattice formed in a monoatomic Fe
layer grown on a Ir(111) surface~\cite{Heinze2011} revealed skyrmions with
diameters as small as a few atom spacings. In addition, it has been
demonstrated that skyrmions can be easily manipulated using spin-polarised
currents of the $10^{6} \,\text{A}\,\text{m}^{-2}$ order~\cite{Jonietz2010, Yu2012}
which is a factor $10^{5}$ to $10^{6}$ smaller than the current
densities required in conventional magneto-electronics. These
unique skyrmion properties point to an opportunity for the realisation of
ambitious novel high-density, power-efficient storage~\cite{Kiselev2011, Fert2013}
and logic~\cite{Zhang2015} devices.

Skyrmionic textures emerge as a consequence of chiral interactions,
also called the Dzyaloshinskii-Moriya Interactions (DMI), that appear
when there is no inversion symmetry in the magnetic system structure.
The lack of inversion symmetry can be either due to a
non-centrosymmetric crystal lattice structure~\cite{Dzyaloshinsky1958, Moriya1960}
in so-called helimagnetic materials, or at interfaces between different materials that
inherently lack inversion symmetry~\cite{Fert1980, Crepieux1998}.
According to this, the Dzyaloshinskii-Moriya interaction can be
classified either as bulk or interfacial, respectively. Skyrmions,
after being predicted~\cite{Bogdanov1989, Bogdanov1999, Rossler2006},
were later experimentally observed in magnetic systems with both
bulk~\cite{Muhlbauer2009, Yu2011, Yu2010, Seki2012, Kanazawa2012} and
interfacial~\cite{Heinze2011, Romming2013} types of DMI.

So far, a major challenge obstructing the development of skyrmion-based
devices has been their thermal and magnetic
stability~\cite{SkyrmionicsEditorial2013}. Only recently, skyrmions were
observed at the room temperature in magnetic systems with
bulk~\cite{Tokunaga2015} and interfacial~\cite{Woo2015, Moreauluchaire2015, Jiang2015}
DMI. However, the magnetic stability of skyrmions in absence of external
magnetic field was reported only for magnetic systems with interfacial DMI in
one-atom layer thin films~\cite{Heinze2011, Sampaio2013}, where the skyrmion
state is stabilised in the presence of magnetocrystalline anisotropy.

The focus of this work is on the zero-field stability of skyrmionic
textures in confined geometries of bulk DMI materials. Zero-field
stability is a crucial requirement for the development of skyrmion-based
devices: devices that require external magnetic fields to be stabilised
are volatile, harder to engineer and consume more
energy. We address the following questions that are relevant for the
skyrmion-based data storage and processing nanotechnology. Can
skyrmionic textures be the ground state (i.e.\ have the lowest energy)
in helimagnetic materials at zero external magnetic field, and if they
can, what is the mechanism responsible for this stability? Do the
demagnetisation energy and magnetisation variation along the out-of-film
direction~\cite{Rybakov2013} have important contribution to the
stability of skyrmionic textures? Is the magnetocrystalline anisotropy
an essential stabilisation mechanism? Are there any other equilibrium
states that emerge in confined helimagnetic nanostructures? How robust
are skyrmionic textures against varying geometry? Do skyrmionic
textures undergo hysteretic behaviour in the presence of an external
magnetic field (crucial for data imprint), and if they do, what is the
skyrmionic texture reversal mechanism?

To resolve these unknowns, we use a full three-dimensional simulation model
that makes no assumption about translational invariance of magnetisation in
the out-of-film direction and takes full account of the demagnetisation
energy. We demonstrate, using this full model, that DMI-induced skyrmionic
textures in confined thin film helimagnetic nanostructures are the lowest
energy states in the absence of both the stabilising external magnetic field and
the magnetocrystalline anisotropy and are able to adapt their size to hosting
nanostructures, providing the robustness for their practical use. We
demonstrate that both the demagnetisation energy and the magnetisation
variation in the out-of-film direction play an important role for the
stability of skyrmionic textures. In addition, we report the parameter
space regions where other magnetisation configurations are in equilibrium.
Moreover, we demonstrate that these zero-field stable
skyrmionic textures undergo hysteretic behaviour when their core orientation
is changed using an external magnetic field, which is crucial for data imprint. The
hysteretic behaviour remains present even in the absence of all relevant
magnetic anisotropies (magnetocrystalline and demagnetisation-based shape
anisotropies), suggesting the existence of a novel Dzyaloshinskii-Moriya-based
shape anisotropy. We conclude the study by showing that the skyrmionic texture
core orientation reversal is facilitated by the Bloch point occurrence and
propagation, where the Bloch point may propagate in either of the two possible
directions. This work is based on the specific cubic helimagnetic material, FeGe
with $70 \,\text{nm}$ helical period, in order to encourage the experimental
verification of our predictions. Other materials could allow either to
reduce the helical period~\cite{Muhlbauer2009, Kanazawa2012} and therefore the hosting
nanostructure size or increase the operating temperature~\cite{Tokunaga2015}.

Some stability properties of DMI-induced isolated
skyrmions in two-dimensional confined systems have been studied
analytically~\cite{Rohart2013, Du2013a, Leonov2013} and using
simulations~\cite{Sampaio2013, Du2013}. However, in all these studies,
either magnetocrystalline anisotropy or an external magnetic field (or both) are
crucial for the stabilisation of skyrmionic textures. In addition,
an alternative approach to the similar problem, in absence of chiral interactions,
where skyrmionic textures can be stabilised at zero external magnetic
field and at room temperature using a strong perpendicular anisotropy, has been studied
analytically~\cite{Guslienko2015}, experimentally~\cite{Buda2001, Moutafis2007},
as well as using simulations~\cite{Moutafis2006}. Our new results,
and in particular the zero-field skyrmionic ground state in isotropic
helimagnetic materials, can only be obtained by allowing the chiral
modulation of magnetisation direction along the film normal, which has
recently been shown to radically change the skyrmion
energetics~\cite{Rybakov2013}.

\section{Results}

\begin{figure*}
\includegraphics{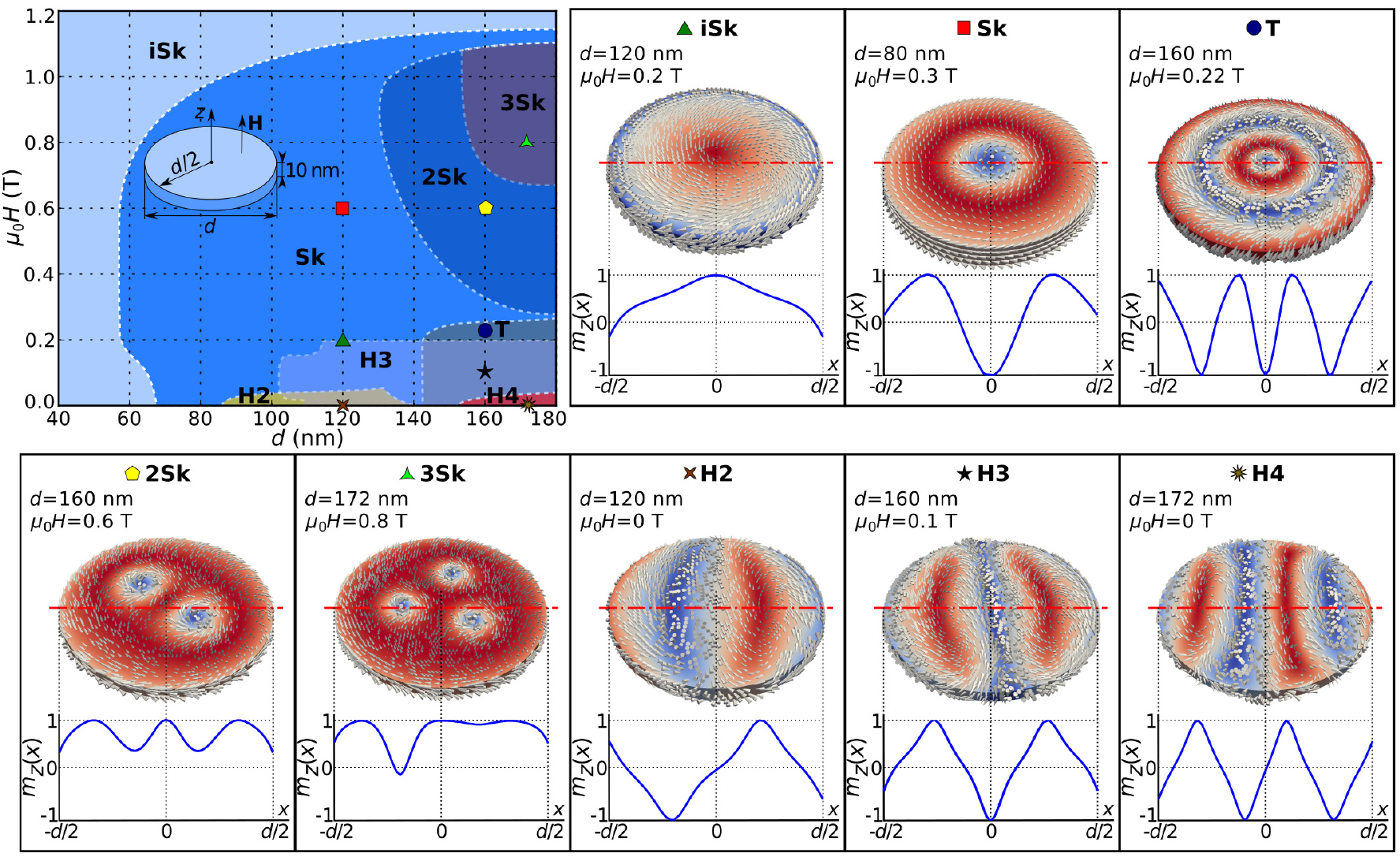}
\caption{\label{fig:equilibrium_states_phase_diagram} \textbf{The
metastability phase diagram and magnetisation configurations of all
identified equilibrium states.} The phase diagram with regions where
different states are in equilibrium together with magnetisation
configurations and out-of-plane magnetisation component $m_{z}(x)$
along the horizontal symmetry line corresponding to different regions
in the phase diagram.}
\end{figure*}

\textbf{Equilibrium states.}\ In order to identify the lowest energy
magnetisation state in confined helimagnetic nanostructures, firstly,
all equilibrium magnetisation states (local energy minima) must be
identified, and secondly, their energies compared. In this section, we
focus on the first step -- identifying the equilibrium magnetisation
states. We compute them by solving a full three-dimensional model
using a finite element based micromagnetic simulator. In particular,
we simulate a thin film helimagnetic FeGe disk nanostructure with
thickness $t = 10 \,\text{nm}$ and diameter $d$, as shown in
Fig.~\ref{fig:equilibrium_states_phase_diagram} inset. The finite
element mesh discretisation is such that the maximum spacing between
two neighbouring mesh nodes is below $3 \,\text{nm}$. The material
parameters are $M_\text{s} = 384 \,\text{kA}\,\text{m}^{-1}$,
$A = 8.78 \,\text{pJ}\,\text{m}^{-1}$, and $D = 1.58 \,\text{mJ}\,\text{m}^{-2}$.
We apply a uniform external magnetic field perpendicular to the
thin film sample, i.e.\ in the positive $z$-direction.
The Methods section contains the details about the model, FeGe material
parameters estimation, as well as the simulator software.

In this section, we determine what magnetisation configurations emerge as the
equilibrium states at different $d$--$H$ parameter space points. In order
to do that, we systematically explore the parameter space by
varying the disk sample diameter $d$ from $40 \,\text{nm}$ to
$180 \,\text{nm}$ and the external magnetic field $\mu_{0}H$ from
$0 \,\text{T}$ to $1.2 \,\text{T}$ in steps of $\Delta d = 4 \,\text{nm}$
and $\mu_{0} \Delta H = 20 \,\text{mT}$, respectively. At every point
in the parameter space, we minimise the energy for a set of different
initial magnetisation configurations: (i) five different skyrmionic
configurations, (ii) three helical-like configurations with different
helical period, (iii) the uniform out-of-plane configuration, and (iv)
three random magnetisation configurations. We use the random
magnetisation configurations in order to capture other equilibrium
states not obtained by relaxing the well-defined initial magnetisation
configurations. The details on how we define and generate initial
magnetisation configurations are provided in the Supplementary Section
S1.

The equilibrium states to which different initial magnetisation configurations
relax in the energy minimisation process (at every $d$--$H$ parameter space point)
we present in the Supplementary Section S2 as a set of ``relaxation diagrams''.
We summarise these relaxation diagrams and determine the phase
space regions where different magnetisation states are in equilibrium, and
show them in Fig.~\ref{fig:equilibrium_states_phase_diagram}. Among the eight
computed equilibrium states, three are radially symmetric and we label them as
iSk, Sk, and T, whereas the other states, marked as H2, H3, H4, 2Sk, and 3Sk,
are not. Subsequently, we discuss the meaning of the chosen labels.

Now, we focus on the analysis of radially symmetric skyrmionic
equilibrium states, supported by computing the skyrmion number
$S$ and scalar value $S_\text{a}$ as defined in the Methods section.
In the first configuration, marked in
Fig.~\ref{fig:equilibrium_states_phase_diagram} as iSk, the out-of-plane
magnetisation component $m_{z}(x)$ profile along the horizontal
symmetry line does not cover the entire $[-1, 1]$ range, as would be the case
for a skyrmion configuration (where the magnetisation vector field $\mathbf{m}$
needs to cover the whole sphere). Accordingly, the scalar value $S_\text{a}$
(Eq.~(\ref{eq:scalar_value_sa_3d}) in the Methods section, and plotted in Supplementary Fig.~2~(b)
for a range of configurations), is smaller than 1.
For these reasons we refer to this skyrmionic equilibrium state as the
incomplete Skyrmion (iSk) state. A similar magnetisation configuration
has been predicted and observed in other works for the case of two-dimensional
systems in the presence of magnetocrystalline anisotropy
where it is called either the quasi-ferromagnetic~\cite{Rohart2013, Sampaio2013}
or edged vortex state~\cite{Du2013, Du2013a}. Because the
iSk equilibrium state clearly differs from the ferromagnetic
configuration and using the word vortex implies the topological charge
of $1/2$, we prefer calling this state the incomplete skyrmion state.
The incomplete Skyrmion (iSk) state emerges as an equilibrium state in
the entire simulated $d$--$H$ parameter space range. In the
second equilibrium state, marked as Sk in
Fig.~\ref{fig:equilibrium_states_phase_diagram}, $m_{z}(x)$ covers
the entire $[-1, 1]$ range, the magnetisation covers
the sphere at least once and, consequently, the skyrmion configuration
is present in the simulated sample. Although the skyrmion number value
(Eq.~(\ref{eq:skyrmion_number_3d}) in the Methods section) for this
solution is $|S| < 1$ due to the additional magnetisation tilting at the disk
boundary~\cite{Rohart2013}, which makes it
indistinguishable from the previously described iSk equilibrium state,
the scalar value is $1 < S_\text{a} < 2$. This state is
referred to as the isolated Skyrmion or just Skyrmion (Sk), in two-dimensional
systems~\cite{Rohart2013, Sampaio2013}, and we use the same name
subsequently in this work. We find that the Sk state is not in
equilibrium for sample diameters smaller than $56 \,\text{nm}$ and
external magnetic field values larger than approximately $1.14 \,\text{T}$.
Finally, the equilibrium magnetisation state marked as T
in Fig.~\ref{fig:equilibrium_states_phase_diagram} covers the sphere
at least twice. In other works, this state together with all other
predicted higher-order solutions (not observed in this work) are
called the ``target states''~\cite{Leonov2013}, and we use
the same Target (T) state name. The analytic model, used for
generating initial states, also predicts the existence of higher-order
target states (Supplementary Fig.~2~(c)). The T magnetisation configuration emerges as an
equilibrium state for samples with diameter $d \geq 144 \,\text{nm}$
and field values $\mu_{0}H \leq 0.24 \,\text{T}$.

The equilibrium states lacking radial symmetry can be classified into
two groups: helical-like (marked as H2, H3, and H4) and multiple
skyrmion (marked as 2Sk and 3Sk) states. The difference between the three
helical-like states is in their helical period. More precisely, in the studied
range of disk sample diameter values, either 2, 3, or 4 helical half-periods,
including the additional magnetisation tilting at the disk
sample edge due to the specific boundary conditions~\cite{Rohart2013},
fit in the sample diameter. Consequently, we refer to these states,
that occur as an equilibrium state for samples larger than $88\,\text{nm}$
and field values lower than $0.2 \,\text{T}$, as H2, H3,
and H4 . The other two radially non-symmetric equilibrium states are
the multiple skyrmion configurations with 2 or 3 skyrmions present in
the sample and we call these equilibrium states 2Sk and 3Sk,
respectively. These configurations emerge as equilibrium states for
samples with $d \geq 132 \,\text{nm}$ and external magnetic field
values between $0.28 \,\text{T} \leq \mu_{0}H \leq 1.06 \,\text{T}$.
\\

\begin{figure}
  \includegraphics{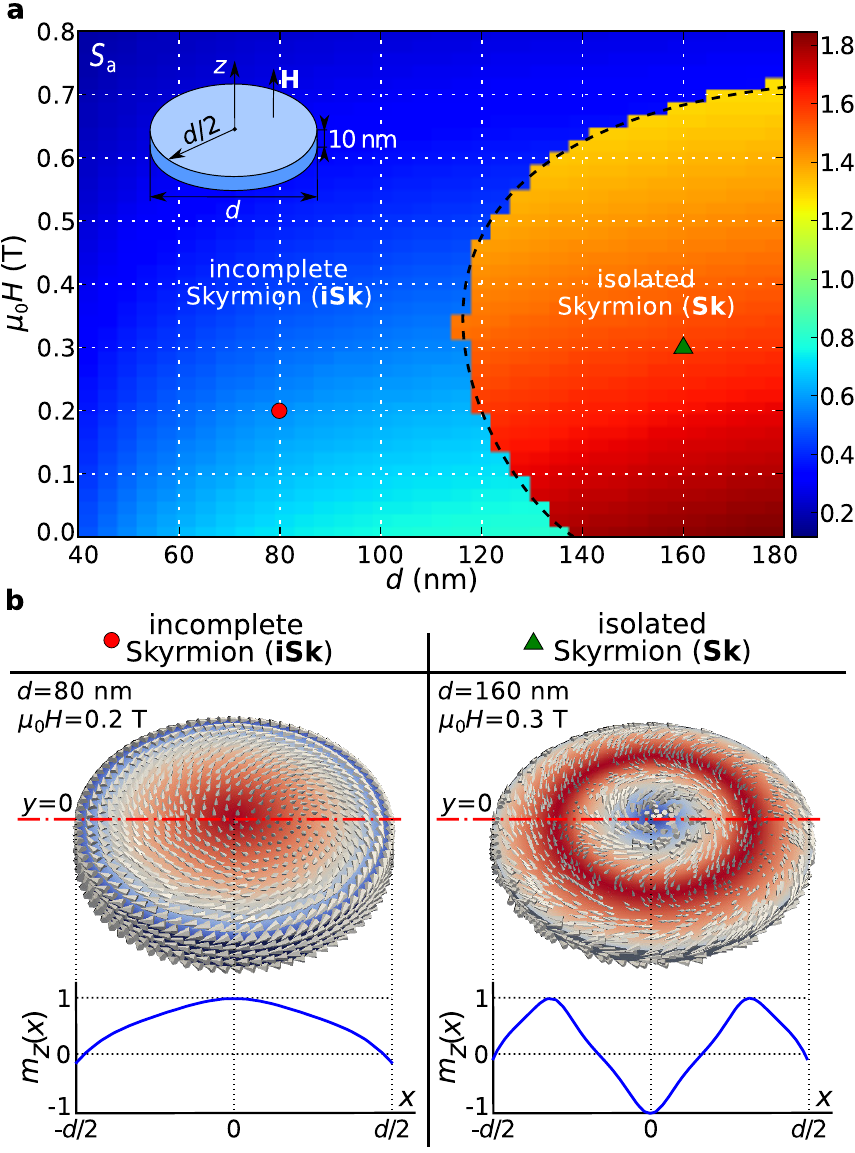}
  \caption{\label{fig:gnd_state_phase_diagram} \textbf{Thin film disk ground
state phase diagram and corresponding magnetisation states.}
(\textbf{a})~The scalar value $S_\text{a}$ for the thin film disk sample with
thickness $t = 10 \,\text{nm}$ as a function of disk diameter $d$ and
external out-of-plane magnetic field $\mathbf{H}$ (as shown in an inset).
(\textbf{b})~Two identified ground states: incomplete Skyrmion (iSk) and isolated 
Skyrmion (Sk) magnetisation configurations at single phase diagram points together
with their out-of-plane magnetisation component $m_{z}(x)$ profiles along
the horizontal symmetry line.}
\end{figure}

\textbf{Ground state.}\ After we identified all observed equilibrium states in confined
helimagnetic nanostructures, in this section we focus on finding the
equilibrium state with the lowest energy at all $d$--$H$ parameter space points.
For every parameter space point ($d$, $H$), after we compute and compare the
energies of all found equilibrium states, we determine the lowest energy
state, and refer to it, in this context, as the ground state. For the
identified ground state, we compute the scalar value $S_\text{a}$ and use it
for plotting a $d$--$H$ phase diagram shown in
Fig.~\ref{fig:gnd_state_phase_diagram}~(a). Discontinuous changes in the
scalar value $S_\text{a}$ define the boundaries between regions where
different magnetisation configurations are the ground state.
In the studied phase space, two different ground states emerge
in the confined helimagnetic FeGe thin film disk samples: one with $S_\text{a}
< 1$ and the other with $1 < S_\text{a} < 2$. The previous discussion of the
$S_\text{a}$ value suggests that these two regions correspond to the
incomplete Skyrmion (iSk) and the isolated Skyrmion (Sk) states. We confirm
this by visually inspecting two identified ground states, taken from the two
phase space points (marked with circle and triangle symbols) in different
regions, and show them in Fig.~\ref{fig:gnd_state_phase_diagram}~(b) together
with their out-of-plane magnetisation component $m_{z}(x)$ along the
horizontal symmetry line.

A key result of this study is that both incomplete Skyrmion (iSk) and isolated
Skyrmion (Sk) are the ground states at zero external magnetic field for
different disk sample diameters. More precisely, iSk is the ground state for
samples with diameter $d < 140 \,\text{nm}$ and Sk is the ground state for 
$d \geq 140 \,\text{nm}$. The Sk changes to the iSk ground state for large values
of external magnetic field.

\begin{figure}
 \includegraphics{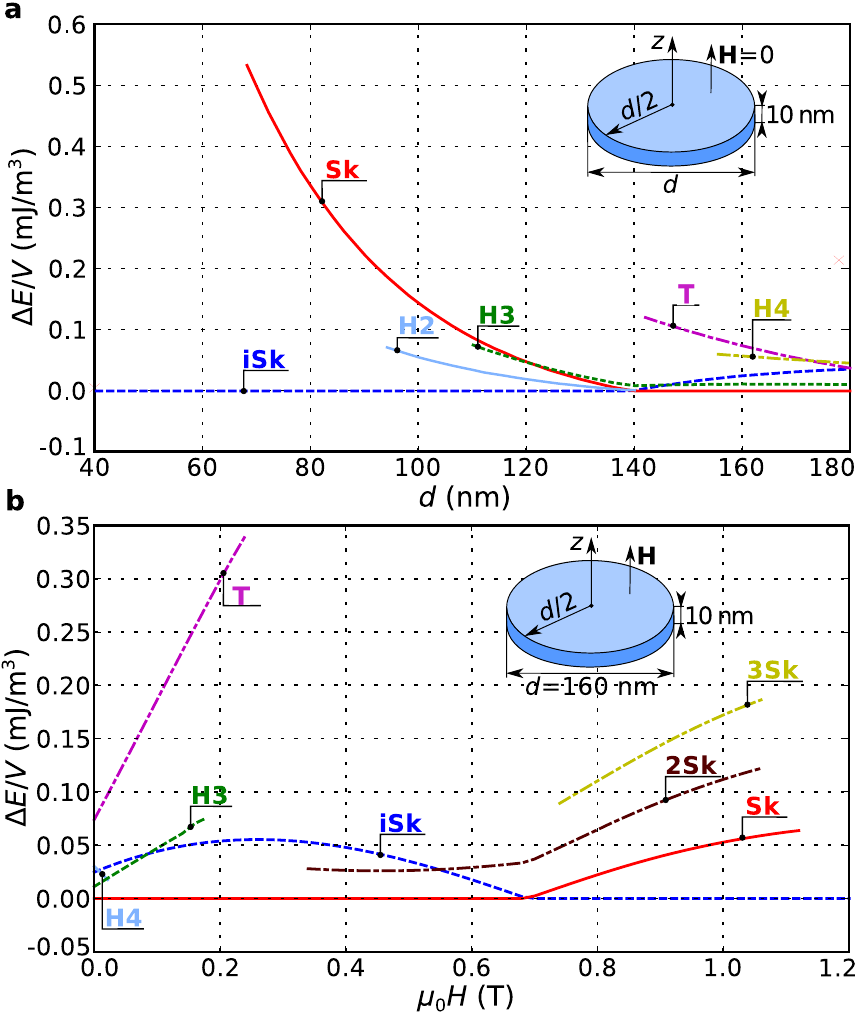}
 \caption{\label{fig:energy_differences} \textbf{The energy density
difference between identified equilibrium states and the
corresponding ground state.} Energy density differences
$\Delta E/V$ at (\textbf{a}) zero field for different sample
diameters $d$ and for (\textbf{b}) sample diameter
$d = 160 \,\text{nm}$ and different external magnetic field values.
Configurations are in equilibrium where the line is shown and
collapse for other diameter or external magnetic field values.}
\end{figure}

The phase diagram in Fig.~\ref{fig:gnd_state_phase_diagram} shows the phase
space regions where iSk and Sk are the ground states, which means that all
other previously identified equilibrium states are metastable. Now, we focus
on computing the energies of metastable states relative to the identified
ground state. Firstly, we compute the energy density $E/V$ for all equilibrium
states, where $E$ is the total energy of the system and $V$ is the disk sample
volume, and then subtract the ground state energy density corresponding to
that phase space point. We show the computed energy density differences
$\Delta E/V$ when the disk sample diameter is changed in steps of
$\Delta d = 2 \,\text{nm}$ at zero external magnetic field in
Fig.~\ref{fig:energy_differences}~(a). Similarly, the case when the disk
sample diameter is $d = 160 \,\text{nm}$ and the external magnetic field is
changed in steps of $\mu_{0} \Delta H = 20 \,\text{mT}$ is shown in
Fig.~\ref{fig:energy_differences}~(b). The magnetisation configurations are
the equilibrium states in the $d$ or $H$ values range where the line is shown
and collapse otherwise.

\begin{figure}
 \includegraphics{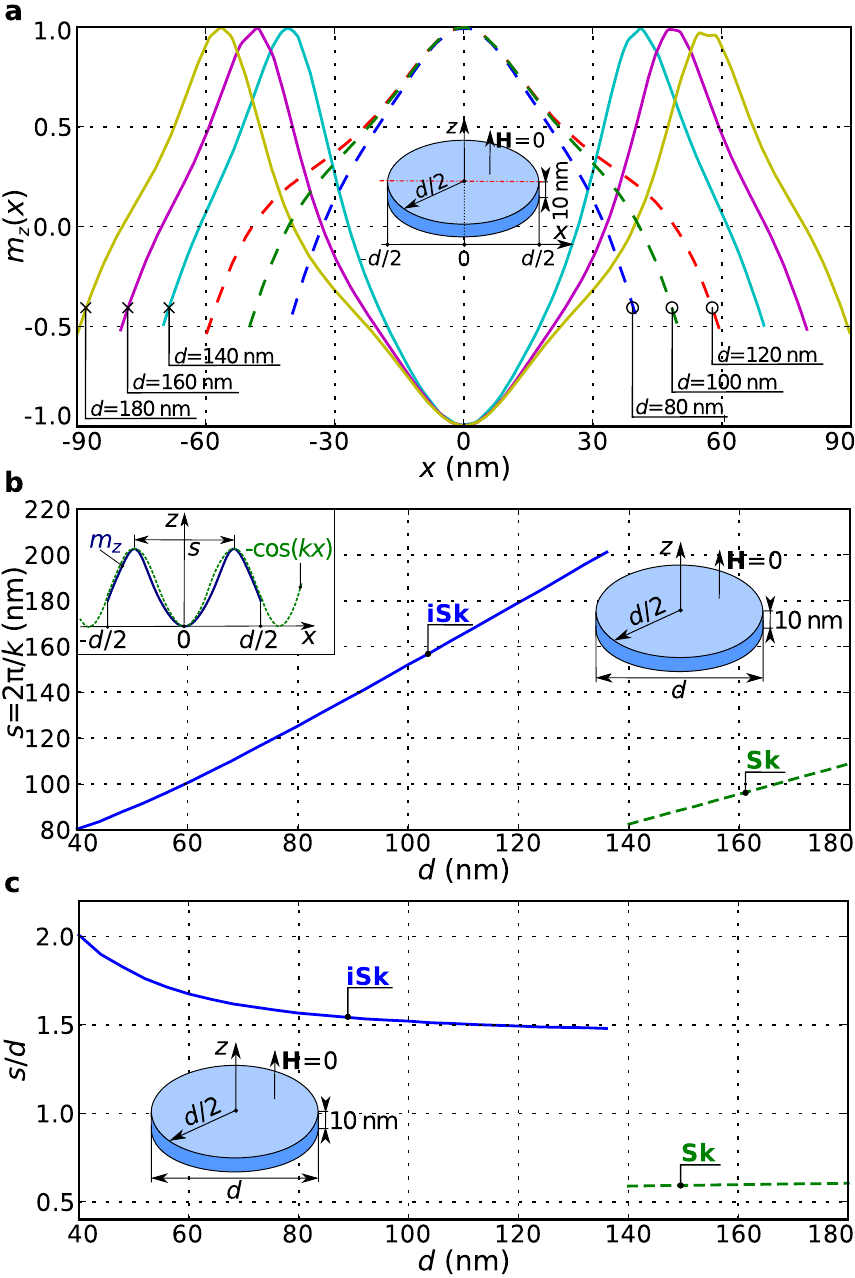}
 \caption{\label{fig:profiles} \textbf{The $m_{z}(x)$ profiles and skyrmionic
texture sizes $s$ for different sizes of hosting nanostructures at zero
external magnetic field.} (\textbf{a})~Profiles of the out-of-plane
magnetisation component $m_{z}(x)$ along the horizontal symmetry line for
different thin film disk sample diameters with thickness $t = 10 \,\text{nm}$
at zero external magnetic field $\mu_{0}H = 0 \,\text{T}$. The curves for
$d \leq 120 \,\text{nm}$ represent incomplete skyrmion ($\circ$) states,
and for $d \geq 140\,\text{nm}$ represent isolated skyrmion ($\times$)
states. (\textbf{b}) The skyrmionic texture size $s = 2\pi/k$ (that can be
interpreted as the length along which the full magnetisation rotation occurs)
as a function of the hosting nanostructure size, obtained by fitting
$m_{z}(x) = \pm \cos(kx)$ to the simulated profile.
(\textbf{c})~The ratio of skyrmionic texture size to disk sample
diameter ($s/d$) as a function of hosting nanostructure size $d$.}
\end{figure}

For the practical use of ground state skyrmionic textures in helimagnetic
nanostructures, their robustness is of great significance due to the
unavoidable variations in the patterning process. Because of that, in
Fig.~\ref{fig:profiles}~(a) we plot the out-of-plane magnetisation component $m_{z}(x)$
along the horizontal symmetry line for the iSk and the Sk ground
state at zero external magnetic field for six different diameters $d$ of the
hosting disk nanostructure: three iSk profiles for $d \leq 120 \,\text{nm}$,
and three Sk profiles for $d \geq 140 \,\text{nm}$. The profiles show that the
two skyrmionic ground states have the opposite core orientations. In the case
of the Sk states, the magnetisation at the core is antiparallel and at the
outskirt parallel to the external magnetic field. This reduces the Zeeman
energy $E_\text{z}= - \mu_0 \int \mathbf{H}\cdot\mathbf{M} \,\text{d}^{3}\mathbf{r}$
because the majority of the magnetisation in the isolated skyrmion
outskirts points in the same direction as the external
magnetic field $\mathbf{H}$. Once the disk diameter is sufficiently small that
less than a complete spin rotation fits into the sample, this orientation is
not energetically favourable anymore and the iSk state emerges. In this iSk
state, the core magnetisation points in the same direction as the external
magnetic field in order to minimise the Zeeman energy. We compute and plot the
skyrmionic texture size $s = 2\pi/k$ as a function of the disk sample diameter
$d$ in Fig.~\ref{fig:profiles}~(b). We obtain the size $s$, that can be interpreted
as the length along which the full magnetisation rotation occurs, by fitting $k$ in
the $f(x) = \pm \cos (kx)$ function to the simulated iSk and Sk $m_{z}(x)$
profiles. In Fig.~\ref{fig:profiles}~(c), we show how the ratio of skyrmionic
texture size to disk sample diameter ($s/d$) depends on the hosting nanostructure size.
Although this ratio is constant ($s/d \approx 0.6$) for the Sk state,
in the iSk case, it is larger for smaller samples and decreases
to $s/d \approx 1.5$ in larger nanostructures. In agreement with
related findings for two-dimensional disk samples~\cite{Du2013a} we find that both
iSk and Sk are able to change their size $s$ in order to accommodate the size
of hosting nanostructure, which provides robustness for the technological use.

\begin{figure*}
  \includegraphics{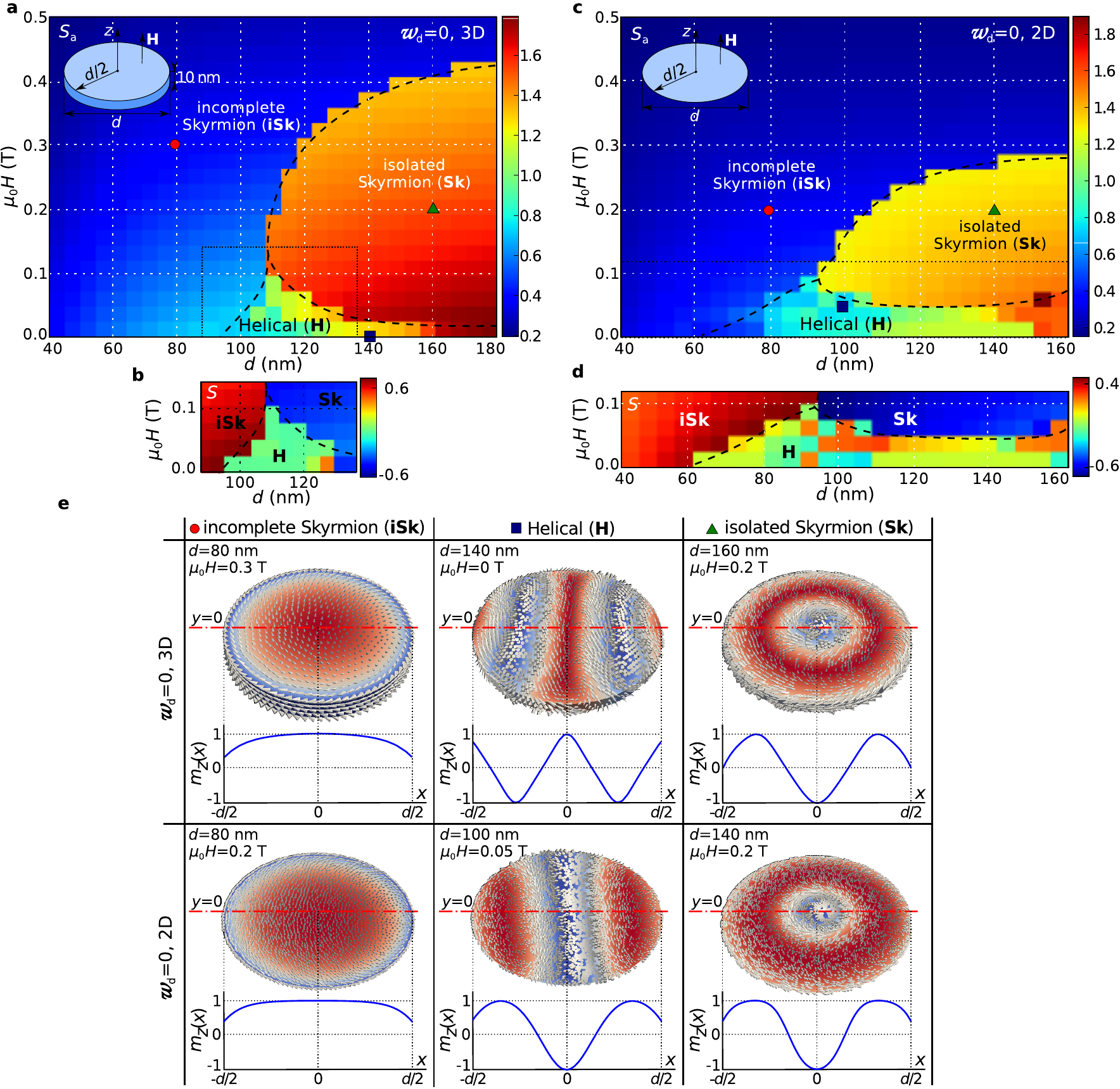}
  \caption{\label{fig:gnd_state_phase_diagram_2nd} \textbf{The ground state
phase diagram in absence of demagnetisation energy contribution.} The scalar
value $S_\text{a}$ as a function of disk sample diameter $d$ and external
magnetic field $H$ computed for the ground state at every phase space
point in absence of demagnetisation energy contribution for (\textbf{a}) a
3d mesh and (\textbf{c}) for a 2d mesh. In order to better resolve the
boundaries of the Helical (H) state region, the skyrmion number $S$ is
shown in (\textbf{b}) and (\textbf{d}). (\textbf{e})~The magnetisation
configurations of three identified ground states as well as the
out-of-plane magnetisation component $m_{z}(x)$ along the horizontal symmetry line.}
\end{figure*}

The emergence of skyrmionic texture ground state in helimagnetic
nanostructures at zero external magnetic field and in absence of
magnetocrystalline anisotropy is unexpected~\cite{SkyrmionicsEditorial2013}.
Now, we discuss the possible mechanisms, apart from the geometrical confinement,
responsible for this stability, in particular (i)~the demagnetisation energy contribution,
and (ii)~the magnetisation variation along the out-of-film direction which can
radically change the skyrmion energetics in infinitely large helimagnetic thin
films~\cite{Rybakov2013}.  We repeat the simulations using the same method and
model as above but ignoring the demagnetisation energy contribution (i.e.\ setting
the demagnetisation energy density $w_\text{d}$ in Eq.~(\ref{eq:energy})
artificially to zero). We then carry out the calculations (i)~on a
three-dimensional (3d) mesh (i.e.\ with spatial resolution in $z$-direction)
and (ii)~on a two-dimensional (2d) mesh (i.e.\ with no spatial resolution
in $z$-direction, and thus not allowing a variation of the magnetisation along the $z$-direction).
The disk sample diameter $d$ is changed between $40 \,\text{nm}$ and $180 \,\text{nm}$
in steps of $\Delta d = 5 \,\text{nm}$ and the external magnetic field
$\mu_{0}H$ is changed systematically between $0 \,\text{T}$ and $0.5 \,\text{T}$
in steps of $\mu_{0}\Delta H = 25 \,\text{mT}$. The two resulting phase diagrams
are shown in Fig.~\ref{fig:gnd_state_phase_diagram_2nd}, where subplots (a) and (c)
show $S_\text{a}$ as a function of $d$ and $H$. Because the scalar value
$S_\text{a}$ does not provide enough contrast to determine the boundaries of the new Helical (H)
ground state region, the skyrmion number $S$ is plotted for the relevant phase
diagram areas and shown in Fig.~\ref{fig:gnd_state_phase_diagram_2nd}~(b) and
Fig.~\ref{fig:gnd_state_phase_diagram_2nd}~(d).

We demonstrate the importance of including demagnetisation effects into the
model by comparing Fig.~\ref{fig:gnd_state_phase_diagram_2nd}~(a) (without
demagnetisation energy) and Fig.~\ref{fig:gnd_state_phase_diagram}~(a)
(with demagnetisation energy). In the absence of the demagnetisation energy,
the isolated Skyrmion (Sk) configuration is not found as the ground state at zero applied
field; instead, Helical (H) configurations have lower energies. At
the same time, the external magnetic field at which the skyrmion configuration
ground state disappears is reduced from about $0.7\,\text{T}$
to about $0.44\,\text{T}$.

By comparing Fig.~\ref{fig:gnd_state_phase_diagram_2nd}~(a)
computed on a 3d mesh and Fig.~\ref{fig:gnd_state_phase_diagram_2nd}~(c)
computed on a 2d mesh, we can see the importance of spatial resolution
in the out-of-plane direction of the thin film, and how it contributes to
the stabilisation of isolated Skyrmion (Sk) state. In the 2d model, the field
range over which skyrmions can be observed as the ground state is further
reduced to approximately [$0.05\,\text{T}$,~$0.28\,\text{T}$].
In the 3d mesh model the Sk configuration can reduce
its energy by twisting the magnetisation at the top of the
disk relative to the bottom of the disk so that along the
$z$-direction the magnetisation starts to exhibit (a part of) the helix that
arises from the competition between symmetric exchange and DMI energy terms,
similar to Ref.~\cite{Rybakov2013}. A similar twist provides
no energetic advantage to the helix configuration, thus the Sk state region
in Fig.~\ref{fig:gnd_state_phase_diagram_2nd}~(a) is
significantly larger than the Sk state region in
Fig.~\ref{fig:gnd_state_phase_diagram_2nd}~(c) where the 2d mesh does not
allow any variation of the magnetisation along the $z$-direction and thus the
partial helix cannot form.

While the isolated Skyrmion (Sk) configuration at zero field is a
metastable state in the absence of demagnetisation energy,
or in 2d models, it is not the ground state anymore as there are
Helical (H) equilibrium configurations that have lower total energy.
The demagnetisation energy appears to suppress these
helical configurations which have a lower energy than the skyrmion. The
variation of the magnetisation along the $z$-direction stabilises the skyrmion
configuration substantially. These findings demonstrate the subtle nature
of competition between symmetric exchange, DM and demagnetisation
interactions, and show that ignoring the demagnetisation energy or
approximating the thin film helimagnetic samples
using two-dimensional models is not generally justified. \\

\textbf{Hysteretic behaviour.}\ The phase diagram in
Fig.~\ref{fig:gnd_state_phase_diagram}~(a) shows the
regions in which incomplete Skyrmion (iSk) and isolated Skyrmion (Sk)
configurations are the ground states. Intuitively, one can assume that for
every sample diameter $d$ at zero external magnetic field, there are two
possible skyrmionic magnetisation configurations of equivalent energy: core
pointing up or core pointing down, suggesting that these textures can be used
for an information bit (0 or 1) encoding. We now investigate this hypothesis
and study whether an external magnetic field can be used to switch the
skyrmionic state orientation (crucial for data imprint) by simulating the
hysteretic behaviour of ground state skyrmionic textures.

\begin{figure*}
 \includegraphics{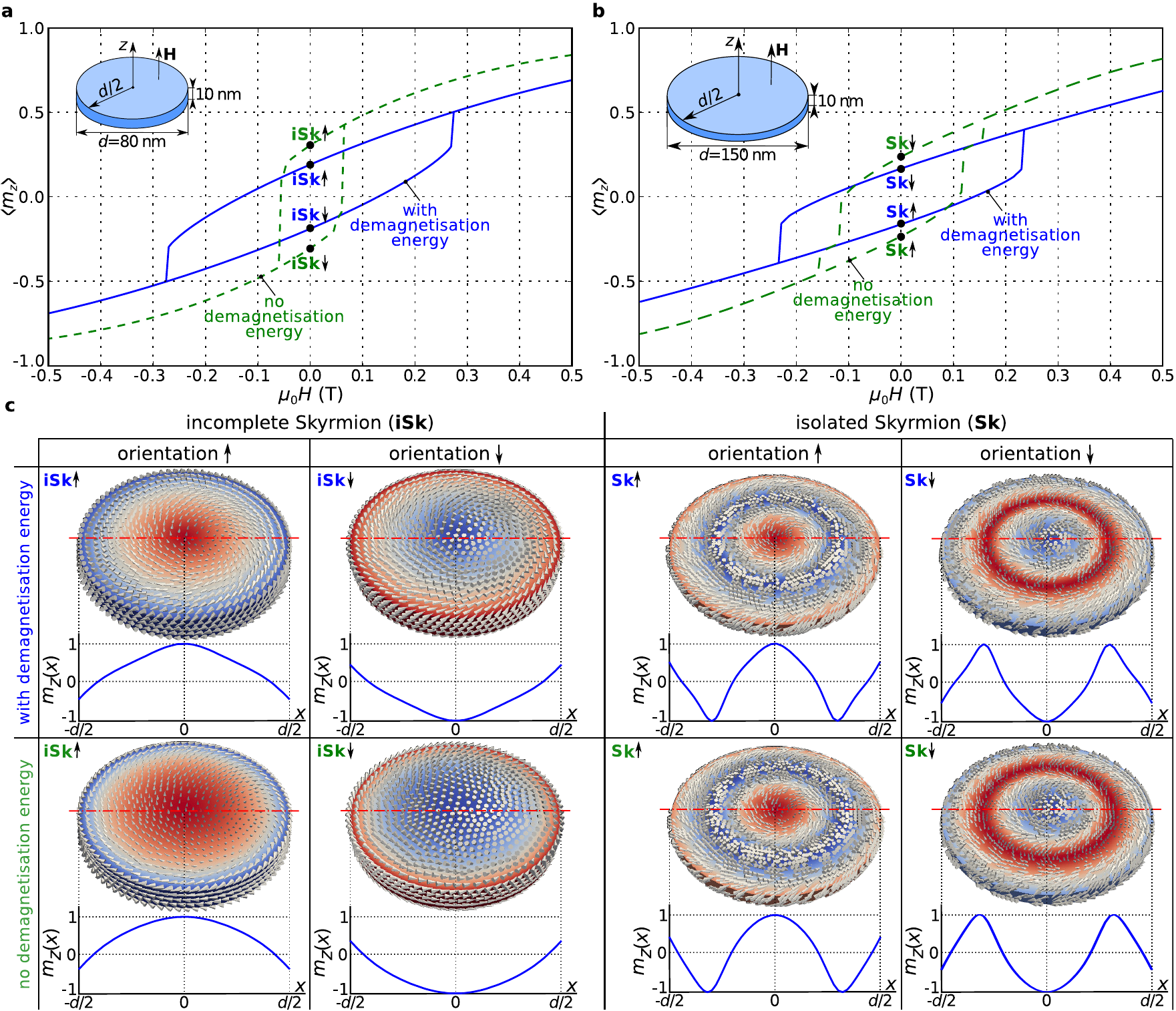}
 \caption{\label{fig:hysteresis} \textbf{Hysteresis loops and obtained
zero-field skyrmionic states with different orientations.} The average
out-of-plane magnetisation component $\langle m_{z} \rangle$ hysteretic
dependence on the external out-of-plane magnetic field $H$ for $10 \,\text{nm}$
thin film disk samples for (\textbf{a})~incomplete Skyrmion (iSk) magnetisation
configuration in $d = 80 \,\text{nm}$ diameter sample and
(\textbf{b})~isolated Skyrmion (Sk) magnetisation configuration in
$d = 150 \,\text{nm}$ diameter sample. (\textbf{c}) The magnetisation states
and $m_{z}(x)$ profiles along the horizontal symmetry lines for positive
and negative iSk and Sk core orientations from $H = 0$ in the hysteresis loop,
both in presence and in absence of demagnetisation energy
(demagnetisation-based shape anisotropy).}
\end{figure*}

We obtain the hysteresis loops in the usual way by evolving the system to an
equilibrium state after changing the external magnetic field, and then using
the resulting state as the starting point for a new evolution. In this way, a
magnetisation loop takes into account the history of the magnetisation
configuration. The external magnetic field $\mu_{0}\mathbf{H}$ is applied in
the positive $z$-direction and changed between $-0.5 \,\text{T}$ and
$0.5 \,\text{T}$ in steps of $\mu_{0}\Delta H = 5 \,\text{mT}$. The hysteresis
loops are represented as the dependence of the average out-of-plane
magnetisation component $\langle m_{z} \rangle$ on the external magnetic field
$H$. The hysteresis loop for a $10 \,\text{nm}$ thin film disk sample with
$d = 80 \,\text{nm}$ diameter in which the incomplete Skyrmion (iSk) is the
ground state is shown in Fig.~\ref{fig:hysteresis}~(a) as a solid line.
Similarly, a solid line in Fig.~\ref{fig:hysteresis}~(b) shows the
corresponding hysteresis loop for a larger disk sample with $d = 150 \,\text{nm}$
diameter in which the isolated Skyrmion (Sk) is the ground state.
The hysteresis between two energetically equivalent skyrmionic magnetisation
states with the opposite core orientation at zero external magnetic field,
shown in Fig.~\ref{fig:hysteresis}~(c), is evident. Moreover, the system does not
relax to any other equilibrium state at any point in the hysteresis loop,
which demonstrates the bistability of skyrmionic textures in studied system.
The area of the open loop in the hysteresis curve is a measure of the work
needed to reverse the core orientation by overcoming the energy barrier
separating the two skyrmionic states with opposite core orientation.

As throughout this work, it is assumed that the simulated helimagnetic
material is isotropic, and thus, the magnetocrystalline anisotropy energy
contribution is neglected. Due to that, one might expect that the obtained hysteresis loops
are the consequence of demagnetisation-based shape anisotropy. To address this,
we simulate hysteresis using the same method, but this time in
absence of the demagnetisation energy contribution. More precisely, the
minimalistic energy model contains only the symmetric exchange and
Dzyaloshinskii-Moriya interactions together with Zeeman coupling to an
external magnetic field. We show the obtained hysteresis loops in
Fig.~\ref{fig:hysteresis}~(a) and (b) as dashed lines. The
hysteretic behaviour remains, although all energy terms that usually 
give rise to the hysteretic behaviour (magnetocrystalline anisotropy and
demagnetisation energies) were neglected. This suggests the existence of a new
magnetic anisotropy that we refer to as the Dzyaloshinskii-Moriya-based shape
anisotropy. \\

\textbf{Reversal mechanism.}\ The hysteresis loops in
Fig.~\ref{fig:hysteresis} show that skyrmionic textures in
confined thin film helimagnetic nanostructures undergo hysteretic
behaviour and that an external magnetic field can be used to change their
orientation from core pointing up to core pointing down and vice versa. In
this section, we discuss the mechanism by which the skyrmionic texture core
orientation reversal occurs. We simulate a $150 \,\text{nm}$ diameter thin
film FeGe disk sample with $t=10 \,\text{nm}$ thickness. The maximum spacing
between two neighbouring finite element mesh nodes is reduced to $1.5 \,\text{nm}$
in order to better resolve the magnetisation field. According to
the hysteresis loop in Fig.~\ref{fig:hysteresis}~(b), the switching field
$H_\text{s}$ of the isolated skyrmion state in this geometry from core
orientation down to core orientation up is $\mu_{0}H_\text{s} \approx -235 \,\text{mT}$.
Therefore, we first relax the system at $-210 \,\text{mT}$
external magnetic field and then decrease it abruptly to $-250 \,\text{mT}$.
We simulate the magnetisation dynamics for $1 \,\text{ns}$, governed by a
dissipative LLG equation~\cite{Gilbert2004} with Gilbert damping
$\alpha = 0.3$~\cite{Sampaio2013}, and record it every $\Delta t = 0.5 \,\text{ps}$.

\begin{figure*}
 \includegraphics{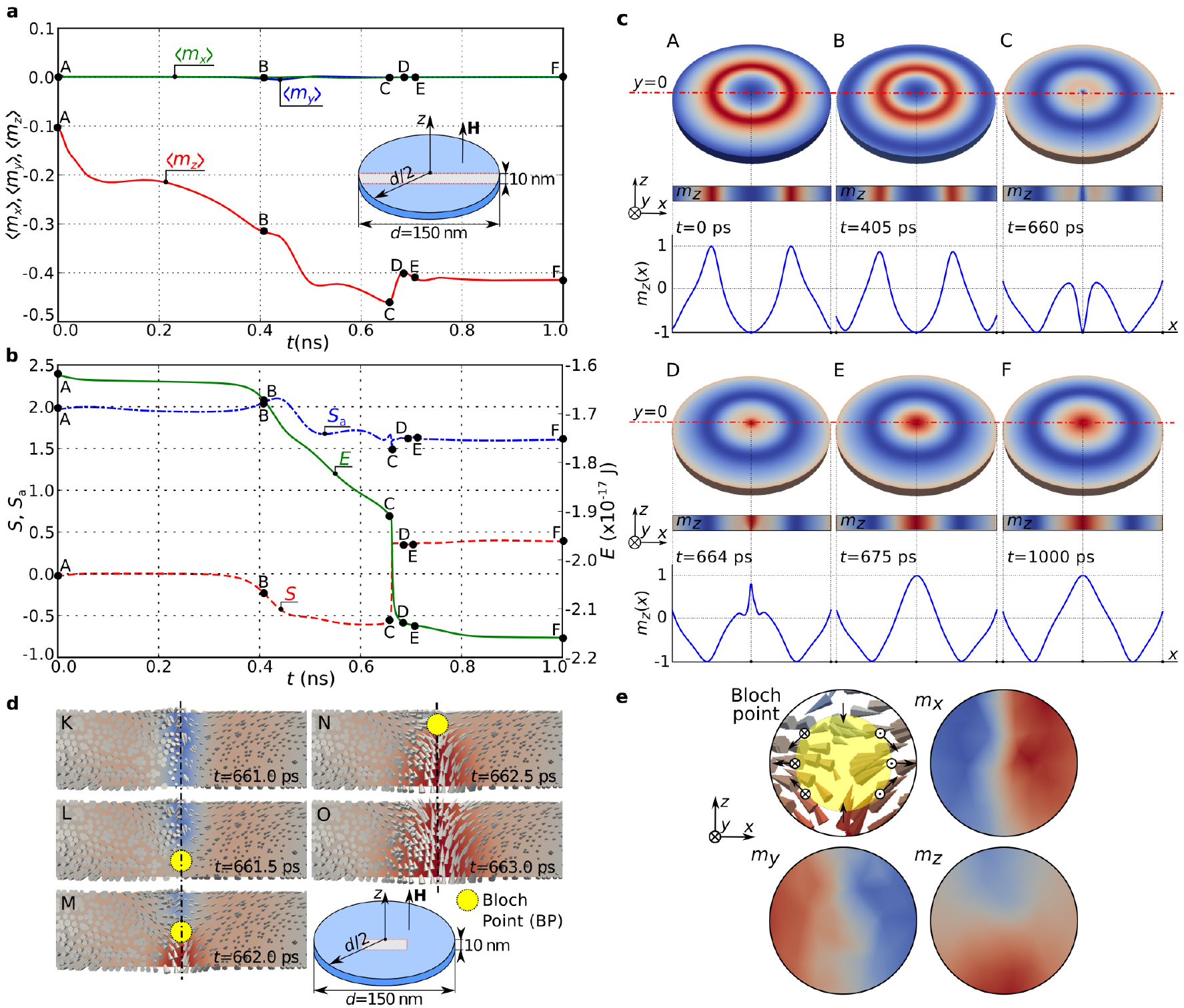}
 \caption{\label{fig:reversal} \textbf{The isolated skyrmion orientation
reversal in confined three-dimensional helimagnetic nanostructure.} (\textbf{a})~The spatially
averaged magnetisation components $\langle m_{x} \rangle$, $\langle m_{y} \rangle$, and
$\langle m_{z} \rangle$ and (\textbf{b})~skyrmion number $S$, scalar value $S_\text{a}$,
and total energy $E$ time evolutions in the reversal process over $1 \,\text{ns}$. The
simulated sample is a $10 \,\text{nm}$ thin film disk with $150 \,\text{nm}$ diameter.
(\textbf{c})~The magnetisation states at different instances of time (points A to F)
together with $m_{z}$ colourmap in the $xz$ cross section and $m_{z}(x)$ profiles along
the horizontal symmetry line. (\textbf{d})~The $m_{z}$ colourmap and magnetisation field
in the central part of $xz$ cross section as shown in an inset together with the position
of Bloch point (BP). (\textbf{e})~The BP structure along with colourmaps of magnetisation
components which shows that the magnetisation covers the closed surface
(sphere surrounding the BP) exactly once.}
\end{figure*}

We now look at how certain magnetisation configuration parameters evolve
during the reversal process. We show the time-dependent average magnetisation
components $\langle m_{x} \rangle$, $\langle m_{y} \rangle$, and
$\langle m_{z} \rangle$ in Fig.~\ref{fig:reversal}~(a), and on the same time axis, the
skyrmion number $S$, scalar value $S_\text{a}$ and total energy $E$ in
Fig.~\ref{fig:reversal}~(b). The initial magnetisation configuration at
$t = 0 \,\text{ns}$ is denoted as A and the final relaxed magnetisation at
$t = 1 \,\text{ns}$ as F. We show in Fig.~\ref{fig:reversal}~(c) the out-of-plane
magnetisation field component $m_{z}$ in the whole sample, in the $xz$ cross
section, as well as along the horizontal symmetry line. At approximately
$662 \,\text{ps}$ the skyrmionic core reversal occurs and
Fig.~\ref{fig:reversal}~(b) shows an abrupt change both in skyrmion number $S$
and total energy $E$. We summarise the reversal process with the help of six
snapshots shown in Fig.~\ref{fig:reversal}~(c). Firstly, in (A-B), the
isolated skyrmion core shrinks. At some point the maximum $m_{z}$ value lowers
from $1$ to approximately $0.1$ (C). After that, the core reverses its
direction (D) and an isolated skyrmion of different orientation is formed (E).
From that time onwards, the core expands in order to accommodate the size of
hosting nanostructure, until the final state (F) is reached. The whole
reversal process is also provided in Supplementary Video 1.

In order to better understand the actual reversal of the skyrmionic texture
core between $t_{1} \approx 661 \,\text{ps}$ and $t_{2} \approx 663 \,\text{ps}$,
we show additional snapshots of the magnetisation vector field
and $m_{z}$ colourmap in the $xz$ cross section in
Fig.~\ref{fig:reversal}~(d). The location marked by a circle in subplots L, M,
and N identifies a Bloch Point (BP): a noncontinuous singularity in the
magnetisation pattern where the magnetisation magnitude vanishes to
zero~\cite{Feldtkeller1965, Doring1968}. Because micromagnetic models assume
constant magnetisation magnitude, the precise magnetisation configuration at
the BP cannot be obtained using micromagnetic simulations~\cite{Andreas2014}.
However, it is known how to identify the signature of the BP in such
situations: the magnetisation direction covers any sufficiently small closed surface
surrounding the BP exactly once~\cite{Slonczewski1975, Thiaville2003}. We illustrate
this property in Fig.~\ref{fig:reversal}~(e) using a vector plot together with
$m_{x}$, $m_{y}$, and $m_{z}$ colour plots that show the structure of a Bloch
point. We conclude that the isolated skyrmion core reversal occurs via Bloch
Point (BP) occurrence and propagation. Firstly, at $t \approx 661.5 \,\text{ps}$
the BP enters the sample at the bottom boundary and propagates
upwards until $t \approx 663 \,\text{ps}$ when it leaves the sample at the top
boundary. In the Supplementary Video 2 the isolated skyrmion core reversal
dynamics is shown.

We note that the Bloch point moves upwards in Fig.~\ref{fig:reversal}~(d) but
one may ask whether an opposite propagation direction can occur and how the
Bloch point structure is going to change. We demonstrate that which of these
two propagation directions will occur in the reversal process depends on the
simulation parameters. The reversal mechanism simulation
was repeated with increased Gilbert damping
($\alpha = 0.35$ instead of $\alpha = 0.3$) and the results showing the downwards
propagation are shown in the Supplementary Section~S3. We hypothesise that
both reversal paths (Bloch point moving upwards or downwards) exhibit the same
energy barriers and that the choice of path is a stochastic process. By
analysing the results from Fig.~\ref{fig:reversal}~(d) and~(e) and
Supplementary Fig.~6, we also observe that the change in the BP propagation
direction implies the change of the BP structure since the out-of-plane
magnetisation component $m_{z}$ field reverses in the vicinity of BP.

\section{Discussion}

Through systematic micromagnetic study of equilibrium states in helimagnetic
confined nanostructures, we identified the ground states and reported the
(meta)stability regions of other equilibrium states. We demonstrated in
Fig.~\ref{fig:gnd_state_phase_diagram} that skyrmionic textures in the form of
incomplete Skyrmion (iSk) and isolated Skyrmion (Sk) configurations are the
ground states in disk nanostructures, and that this occurs in a wide
$d$--$H$ parameter space range. We have carried out similar studies for a
square geometry and obtain qualitatively similar results. Of particular
importance is that iSk and Sk states are the ground states at zero external
magnetic field which is in contrast to infinite thin film and bulk
helimagnetic samples. We note that neither an external magnetic field is
necessary nor magnetocrystalline anisotropy is required for this stability. We
also note in Fig.~\ref{fig:profiles}~(c) that there is significant flexibility
in the skyrmionic texture size which provides robustness for technology built
on skyrmions, where fabrication of nanostructures and devices introduces
unavoidable variation in geometries.

We have established that including the demagnetisation interaction is crucial
for the system investigated here, i.e.\ in the absence of demagnetisation
effects, there are other magnetisation configurations with energies lower than
that of the incomplete and isolated skyrmion. We also note that the
translational variance of the magnetisation from the lower side of the thin film
(at $z = 0 \,\text{nm}$) to the top (at $z = 10 \,\text{nm}$) is essential for the
physics reported here: if we use a two-dimensional micromagnetic simulation
(i.e.\ assuming translational invariance of the magnetisation $\mathbf{m}$ in
the out-of-plane direction), the isolated skyrmion configuration does not
arise as the ground state. Our interpretation is that for skyrmion-like
configurations the twist of $\mathbf{m}$ between top and bottom layer allows
the system's energy to reduce significantly while such a reduction is less
beneficial for other configurations such as helices; inline with recent
predictions in the case of infinite thin films~\cite{Rybakov2013}.
Accordingly, we conclude that three-dimensional
helimagnetic nanostructure models, where demagnetisation energy
contribution is neglected, or the geometry approximated using a two-dimensional
mesh, are not generally justified.

Because of the specific boundary conditions~\cite{Rohart2013} and
the importance of including the demagnetisation energy contribution, our
predictions cannot be directly applied to other helimagnetic materials 
without repeating the stability study. For instance, although the size of
skyrmionic textures in this study was based on cubic FeGe
helimagnetic material with helical period $L_\text{D} = 70 \,\text{nm}$,
in order to encourage the experimental verification of our predictions, this
study could be repeated for materials with smaller $L_\text{D}$. In such
materials the skyrmionic core size is considerably reduced, which allows the
reduction of hosting nanostructure size and is an essential requirement for
advancing future information storage technologies. Similarly,
the ordering temperature of simulated FeGe helimagnetic material,
$T_\text{C} = 278.7 \,\text{K}$~\cite{Lebech1989}, is lower than the room temperature,
which means that a device operating at the room temperature cannot be
constructed using this material. Because of that, in Supplementary Section S4,
we demonstrate that our predictions are still valid if the ordering temperature
of simulated B20 helimagnetic material is artificially
increased to $350 \,\text{K}$.

We demonstrate in Fig.~\ref{fig:hysteresis} that skyrmionic textures in
confined helimagnetic nanostructures exhibit hysteretic behaviour as a consequence of energy
barriers between energetically equivalent stable configurations (skyrmionic
texture core pointing up or down). In the absence of magnetocrystalline
anisotropy and if the demagnetisation energy (demagnetisation-based shape
anisotropy) is removed from the system's Hamiltonian, the hysteretic behaviour
is still present, demonstrating the existence of a novel
Dzyaloshinskii-Moriya-based shape anisotropy.

Finally, we show how the reversal of the isolated skyrmion core orientation is
facilitated by the Bloch point occurrence and propagation, and demonstrate
that the Bloch point can propagate in both directions along the out-of-plane
$z$-direction.

All data obtained by micromagnetic simulations in this study and used to
create figures both in the main text and in the Supplementary Information
are included in Supplementary Data.

\section{Methods}

\textbf{Model.}\ We use an energy model consistent with a non-centrosymmetric cubic B20
(P2$_{1}$3 space group) crystal structure. This is appropriate for a range of
isostructural compounds and pseudo-binary alloys in which skyrmionic textures
have been experimentally
observed~\cite{Muhlbauer2009, Jonietz2010, Yu2010, Yu2011, Yu2012, Wilhelm2012, Huang2012, Seki2012}.
The magnetic free energy of the system $E$ contains several contributions
and can be written in the form:
\begin{equation}
  \label{eq:energy}
  E = \int \left[ w_\text{ex} + w_\text{dmi} + w_\text{z} + w_\text{d} + w_\text{a} \right] \,\text{d}^{3}r.
\end{equation}
The first term is the symmetric exchange energy density
$w_\text{ex} = A \left[ (\nabla m_{x})^{2} + (\nabla m_{y})^{2} + (\nabla m_{z})^{2} \right]$
with exchange stiffness material parameter $A$, where $m_{x}$, $m_{y}$, and
$m_{z}$ are the Cartesian components of the vector $\mathbf{m} = \mathbf{M}/M_\text{s}$
that describes the magnetisation~$\mathbf{M}$, with
$M_\text{s}=|\mathbf{M}|$ being the saturation magnetisation. The second term
is the Dzyaloshinskii-Moriya Interaction (DMI) energy density
$w_\text{dmi} = D \mathbf{m} \cdot \left( \nabla \times \mathbf{m} \right)$, obtained by
constructing the allowed Lifshitz invariants for the crystallographic class
T~\cite{Bak1980, Bogdanov1989}, where $D$ is the material parameter. The third
term is the Zeeman energy density term $w_\text{z} = - \mu_{0}\mathbf{H} \cdot \mathbf{M}$
which defines the coupling of magnetisation to an external
magnetic field~$\mathbf{H}$. The $w_\text{d}$ term represents the
demagnetisation (magnetostatic) energy density. The last term
$w_\text{a}$ is the magnetocrystalline anisotropy energy density, and because
the simulated material is assumed to be isotropic, we neglect it throughout
this work. Neglecting this term also allows us to determine whether the
magnetocrystalline anisotropy is a crucial mechanism allowing the stability of
skyrmionic textures in confined helimagnetic nanostructures.

The Landau-Lifshitz-Gilbert (LLG) equation~\cite{Gilbert2004}:
\begin{equation}
\label{eq:LLG}
  \frac{\partial \mathbf{m}}{\partial t} = \gamma^{*} \mathbf{m} \times \mathbf{H}_\text{eff} + \alpha\mathbf{m} \times \frac{\partial \mathbf{m}}{\partial t},
\end{equation}
governs the magnetisation dynamics, where $\gamma^{*} = \gamma (1 + \alpha^{2})$,
with $\gamma < 0$ and $\alpha$ being the gyromagnetic ratio
and Gilbert damping, respectively. We compute the effective
magnetic field using 
$\mathbf{H}_\text{eff} = -(\delta w / \delta \mathbf{m}) / (\mu_{0}M_\text{s})$,
where $w$ is the total energy density functional. With this model, we solve
for magnetic configurations $\mathbf{m}$ using the
condition of minimum torque arrived by integrating a set of dissipative, 
time-dependent equations. We validated the boundary
conditions by a series of simulations reproducing the results in
Ref.~\cite{Rohart2013, Sampaio2013}.
\\

\textbf{Simulator.}\ We developed a micromagnetic simulation software, inspired
by the Nmag simulation tool~\cite{Fischbacher2007, Nmag}. Unlike Nmag, we use the
FEniCS project~\cite{Logg2012} instead of the Nsim multi-physics
library~\cite{Fischbacher2007} for the finite element low-level operations.
In addition, we use IPython~\cite{Perez2007, IPython} and Matplotlib~\cite{Hunter2007, Matplotlib}
extensively in this work. \\

\textbf{Material parameters.}\ We estimate the material parameters
in our simulations to represent the cubic B20 FeGe helimagnet 
with four Fe and four Ge atoms per unit
cell~\cite{Pauling1948} and crystal lattice constant
$a = 4.7 \,\text{\AA}$~\cite{Richardson1967}. The local magnetic moments of iron and
germanium atoms are $1.16 \mu_\text{B}$ and $-0.086
\mu_\text{B}$~\cite{Yamada2003}, respectively, where $\mu_\text{B}$ is the
Bohr magneton constant. Accordingly, we estimate the saturation magnetisation
as $M_\text{s} = 4N(1.16 - 0.086)\mu_\text{B} = 384 \,\text{kA}\,\text{m}^{-1}$,
with $N = a^{-3}$ being the number of lattice unit cells in a cubic metre.
The spin-wave stiffness is $D_\text{sw} = a^{2}T_\text{C}$~\cite{Grigoriev2005},
where the FeGe ordering temperature is $T_\text{C} = 278.7 \,\text{K}$~\cite{Lebech1989}.
Consequently, the exchange stiffness parameter value is
$A = D_\text{sw}M_\text{s}/(2g\mu_\text{B}) = 8.78 \,\text{pJ}\,\text{m}^{-1}$~\cite{Hamrle2009},
where $g \approx 2$ is the Land\'{e} $g$-factor. The estimated DMI material
parameter $D$ from the long-range FeGe helical period
$L_\text{D} = 70 \,\text{nm}$~\cite{Lebech1989},
using $L_\text{D} = 4\pi A/|D|$~\cite{Wilhelm2012}, is $|D| = 1.58 \,\text{mJ}\,\text{m}^{-2}$. \\

\textbf{Skyrmion number $S$ and injective scalar value $S_\mathrm{a}$.}\ In order
to support the discussion of skyrmionic textures, the topological
skyrmion number~\cite{Heinze2011}
\begin{equation}
  \label{eq:skyrmion_number_2d}
  S^\text{2D} = \frac{1}{4\pi} \int \mathbf{m} \cdot \left( \frac{\partial \mathbf{m}}{\partial x} \times \frac{\partial \mathbf{m}}{\partial y}\right) \,\text{d}^{2}r,
\end{equation}
can be computed for two-dimensional samples hosting the
magnetisation configuration. However, for confined systems, the skyrmion number $S^\text{2D}$ is not
quantised into integers~\cite{Sampaio2013, Du2013}, and therefore, a more
suitable name for $S^\text{2D}$ may be the ``scalar spin chirality'' (and consequently
the expression under an integral would be called the ``spin chirality
density''), but we will follow the existing literature~\cite{Sampaio2013, Du2013}
and refer to $S^\text{2D}$ as the skyrmion number. We show its dependence on
different skyrmionic textures that can be observed in confined helimagnetic nanostructures
in Supplementary Fig.~2~(b), demonstrating that the skyrmion number in confined
geometries is not an injective function since it does not preserve
distinctness (one-to-one mapping between skyrmionic textures and skyrmion
number value $S^\text{2D}$). Therefore, for two-dimensional samples, we define a different scalar value
\begin{equation}
  \label{eq:scalar_value_sa_2d}
  S^\text{2D}_\text{a} = \frac{1}{4\pi} \int \left|\mathbf{m} \cdot \left( \frac{\partial \mathbf{m}}{\partial x} \times \frac{\partial \mathbf{m}}{\partial y}\right) \right| \,\text{d}^{2}r,
\end{equation}
and show its dependence on different skyrmionic textures in Supplementary
Fig.~2~(b). This scalar value is injective and provides
necessary distinctness between $S_\text{a}^{2D}$ values for different skyrmionic
states. In terms of the terminology discussion above regarding $S^\text{2D}$, the entity
$S_\text{a}^{2D}$ describes the ``scalar absolute spin chirality''. We also
emphasise that although the skyrmion number $S^\text{2D}$ has a clear
mathematical~\cite{Braun2012} and physical~\cite{Schulz2012} interpretation, we define
the artificial injective scalar value $S_\text{a}$ only to support the
classification and discussion of different skyrmionic textures observed
in this work.

Skyrmion number $S^\text{2D}$ and artificially defined scalar
value $S_\text{a}^\text{2D}$, given by Eq.~(\ref{eq:skyrmion_number_2d}) and
Eq.~(\ref{eq:scalar_value_sa_2d}), respectively, are valid only for the
two-dimensional samples hosting the magnetisation configuration.
However, in this work, we also study three-dimensional samples and,
because of that, we now define a new set of expressions taking into
account the third dimension. The skyrmion number in three-dimensional
samples $S^\text{3D}$ we compute using
\begin{equation}
  \label{eq:skyrmion_number_3d}
  S^\text{3D} = \frac{1}{8\pi} \int \mathbf{m} \cdot \left( \frac{\partial \mathbf{m}}{\partial x} \times \frac{\partial \mathbf{m}}{\partial y}\right) \,\text{d}^{3}r,
\end{equation}
as suggested by Lee et al.~\cite{Lee2009}, which results in a value
proportional to the anomalous Hall conductivity. Similar to the
two-dimensional case, we also define the artificial injective
scalar value $S_\text{a}^\text{3D}$ for three-dimensional samples as
\begin{equation}
  \label{eq:scalar_value_sa_3d}
  S^\text{3D}_\text{a} = \frac{1}{8\pi} \int \left|\mathbf{m} \cdot \left( \frac{\partial \mathbf{m}}{\partial x} \times \frac{\partial \mathbf{m}}{\partial y}\right) \right| \,\text{d}^{3}r.
\end{equation}
In order to allow the $S^\text{3D}_\text{a}$ value to fall within
the two-dimensional skyrmionic textures classification scheme,
we normalise the computed $S^\text{3D}_\text{a}$ value by a constant ($t/2$,
where $t$ is the sample thickness).

For simplicity, in this work, we refer to both
two-dimensional and three-dimensional skyrmion number
and scalar value expressions as $S$ and $S_\text{a}$ because it is
always clear what expression has been used according to the
dimensionality of the sample.

\section{References}

%\bibliographystyle{naturemag}
%\bibliography{../bibliography/library}

\section{Acknowledgements}

This work was financially supported by the EPSRC’s Doctoral
Training Centre (DTC) grant EP/ G03690X/1. R.L.S. acknowledges
the EPSRC’s EP/M024423/1 grant support. D.C.-O. acknowledges
the financial support from CONICYT Chilean scholarship programme
Becas Chile (72140061). We acknowledge the use of the IRIDIS High
Performance Computing Facility, and associated support services
at the University of Southampton, in the completion of this work.
We also thank Karin Everschor-Sitte for helpful discussions.

\section{Author Contributions}

M.B. and H.F conceived the study, and M.B. performed micromagnetic
simulations. R.L.S. devised the analytic model and discussed its implications.
R.C. contributed to the simulations and analysis of equilibrium states. D.C.,
M.-A.B., M.A., W.W., M.B., R.C., M.V., D.C.-O. and H.F. developed the
micromagnetic finite element based simulator. M.V. and M.A. enabled running
simulations on IRIDIS High Performance Computing Facility. M.B., H.F., R.L.S.,
and O.H. interpreted the data and prepared the manuscript.

\section{Competing financial interests}

The authors declare no competing financial interests.

\clearpage


\section*{Supplementary material}

# Supplementary Information: Ground state search, hysteretic behaviour, and reversal mechanism of skyrmionic textures in helimagnetic nanostructures

Marijan Beg,[1, *] Rebecca Carey,[1] Weiwei Wang,[1] David Cortés-Ortuño,[1] Mark Vousden,[1] Marc-Antonio Bisotti,[1]
Maximilian Albert,[1] Dmitri Chernyshenko,[1] Ondrej Hovorka,[1] Robert L. Stamps,[2] and Hans Fangohr[1, †]

[1]*Faculty of Engineering and the Environment, University of Southampton, Southampton SO17 1BJ, United Kingdom*
[2]*SUPA School of Physics and Astronomy, University of Glasgow, Glasgow G12 8QQ, United Kingdom*

## SUPPLEMENTARY SECTION S1: INITIAL MAGNETISATION CONFIGURATIONS

The magnetisation configurations that we use as initial states for the full three-dimensional micromagnetic simulations are shown in Supplementary Fig. 1. For every point in the $d$–$H$ parameter space, we relax twelve different initial magnetisation configurations. These are the five skyrmionic configurations (A, B, C, D, and E), three helical configurations (H2, H3, and H4), the uniform configuration (U), and three random magnetisation configurations (R).

Now, we introduce an approximate analytic model that helps us generate a range of physically meaningful and reproducible initial skyrmionic magnetisation configurations labelled A-E in Supplementary Fig. 1. The used DMI energy density term (see Methods section of the main text) is consistent with the helimagnetic material of crystallographic class T, and one expects a skyrmionic texture configuration with no radial spin component (chiral skyrmion) to emerge [1]. Consequently, if we consider a two-dimensional disk sample of radius $R$ in the plane containing the $x$ and $y$ axes, as shown in Supplementary Fig. 2 (b) inset, an approximation of the chiral skyrmionic magnetisation texture (for $D > 0$), in cylindrical coordinates $(\rho, \varphi, z)$, is

$$
\begin{aligned}
m_\rho &= 0, \\
m_\varphi &= \sin(k\rho), \\
m_z &= -\cos(k\rho),
\end{aligned}
\tag{1}
$$

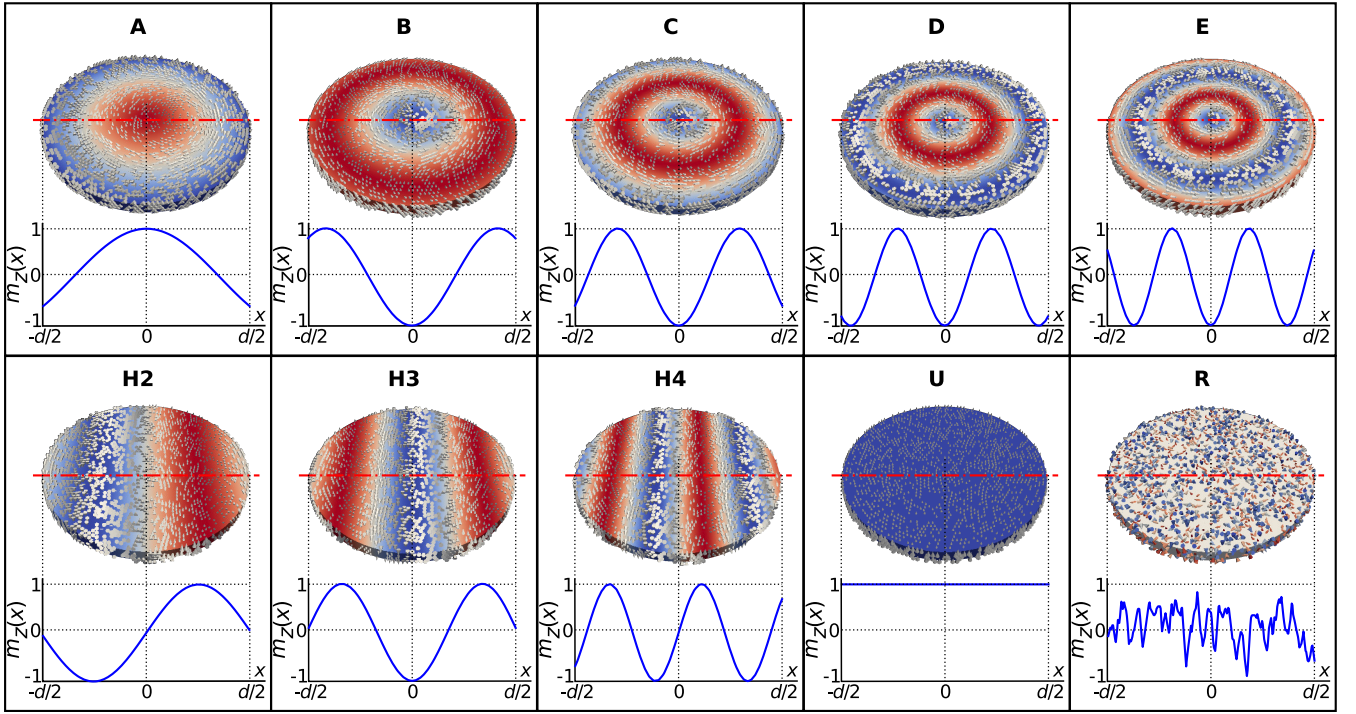

FIG. 1. **The magnetisation configurations relaxed using the full three-dimensional micromagnetic simulations.** The configurations labelled A-E correspond to the first five solutions of the approximate analytic model, whereas the three helical states with 2, 3, and 4 helical half-periods are labelled as H2, H3, and H4, respectively. The uniform out-of-plane magnetisation state is labelled as U, and an example of random magnetisation state is marked with R.

where $k = 2\pi/s$ is a measure of the skyrmionic texture size $s$.

An equilibrium configuration requires that the torque exerted on the magnetisation $\mathbf{m}$ vanishes at every point in sample ($\mathbf{m} \times \mathbf{H}_{\text{eff}} = 0$), including the boundary. The effective field functional $\mathbf{H}_{\text{eff}} = -(\delta w/\delta \mathbf{m})/(\mu_0 M_{\text{s}})$, due to only symmetric exchange and DMI energy contributions, in absence of an external magnetic field is

$$\mathbf{H}_{\text{eff}} = \frac{2}{\mu_0 M_{\text{s}}} \left[ A\nabla^2 \mathbf{m} - D\left(\nabla \times \mathbf{m}\right) \right],$$

(2)

for any magnetisation texture. Computing this expression for the radially symmetric approximate skyrmionic texture model $\mathbf{m} = \sin(k\rho)\hat{\boldsymbol{\varphi}} - \cos(k\rho)\hat{\mathbf{z}}$ results in

$$\begin{aligned}
\mathbf{H}_{\text{eff}} = &\frac{2}{\mu_0 M_{\text{s}}} \left[ \left( Dk - Ak^2 - \frac{A}{\rho^2} \right) \sin(k\rho) + \frac{Ak}{\rho} \cos(k\rho) \right] \hat{\boldsymbol{\varphi}} + \\
&\frac{2}{\mu_0 M_{\text{s}}} \left[ \left( \frac{Ak}{\rho} - \frac{D}{\rho} \right) \sin(k\rho) + \left( Ak^2 - Dk \right) \cos(k\rho) \right] \hat{\mathbf{z}}.
\end{aligned}$$

(3)

Consequently, the torque exerted on the magnetisation $\mathbf{m}$ is

$$\mathbf{m} \times \mathbf{H}_{\text{eff}} = \frac{2}{\mu_0 M_{\text{s}}} \left[ \frac{Ak}{\rho} - \frac{D}{\rho} \sin^2(k\rho) - \frac{A}{2\rho^2} \sin(2k\rho) \right] \hat{\mathbf{r}}.$$

(4)

Requiring the torque to vanish at the disk boundary $\rho = R$ results in the zero-torque condition:

$$g(kR) \equiv -P \sin^2(kR) - \frac{\sin(2kR)}{2kR} + 1 = 0,$$

(5)

where $P = D/kA$. The analysis of this condition shows that the parameter $P$ must satisfy the inequality $P > 2/3$ in order for $g(kR)$ to have roots and, thus, a skyrmionic texture core to exist in at least metastable equilibrium. In Supplementary Fig. 2 (a), we plot the zero-torque condition as a function of $kR$ for different values of $P$. Since $P = 2/3$ is the boundary case (dotted line in the plot), all plots for $P < 2/3$ have no solutions, whereas the zero-torque condition has multiple solutions if $P > 2/3$.

The skyrmionic magnetisation configurations used as initial states in the energy minimisation process, corresponding to the zero-torque condition solutions (marked A-F) for $P = 2$, we show in Fig. 2 (c). In order to support the discussion of these magnetisation configurations, we compute the skyrmion number $S$, which for our approximate analytic model results in $S = (\cos(kR) - 1)/2$. Its dependence on $kR$, presented in Fig. 2 (b) as a dashed line, shows that the skyrmion number for the skyrmionic textures in confined nanostructures is not an injective function since it does not preserve distinctness (one-to-one mapping between $kR$ and skyrmion number value $S$). Therefore, a different scalar value $S_{\text{a}}$ is computed and from its dependence on $kR$, shown in Fig. 2 (b) as a solid line, we conclude that this scalar value is injective and provides necessary distinctness between $S_{\text{a}}$ values for different skyrmionic states. The details about skyrmion number $S$ and scalar value $S_{\text{a}}$ are in the Methods section of the main text.

For the helical state, which emerges as a consequence of the Dzyaloshinskii-Moriya interaction considered in this work, we expect that the magnetisation vector at any point is perpendicular to the helical propagation direction (Bloch-wall-like configuration). Consequently, if both $x$ and $y$ axis are in the plane of the thin-film sample and the $x$ axis is chosen as a propagation direction, the helical magnetisation configuration in Cartesian coordinates is

$$\begin{aligned}
m_x &= 0, \\
m_y &= \cos(k_{\text{h}}x), \\
m_z &= \sin(k_{\text{h}}x),
\end{aligned}$$

(6)

where $k_{\text{h}} = 2\pi/\lambda$, with $\lambda$ being the helical period [2].

Now, we investigate whether the helical period $\lambda$ in confined nanostructures is independent on the sample diameter, and if not, what are the helical period values that can occur in our simulated system. After relaxing helical configurations and varying both the helical period and disk sample diameter (up to $180\,\text{nm}$), we find that all relax to a limited set of helical states with different helical periods. More precisely, the observed relaxed helical states consist of either 2, 3, or 4 helical half-periods along the disk sample diameter, including the characteristic magnetisation tilting at the boundary [3]. Thus, we define three different helical magnetisation configurations as initial states with helical periods $2d/2$, $2d/3$, or $2d/4$, where $d$ is the disk sample diameter, and are named H2, H3, and H4, respectively. Magnetisation configurations of these states, together with their $m_z(x)$ profiles along the horizontal symmetry, are shown in Supplementary Fig. 1.

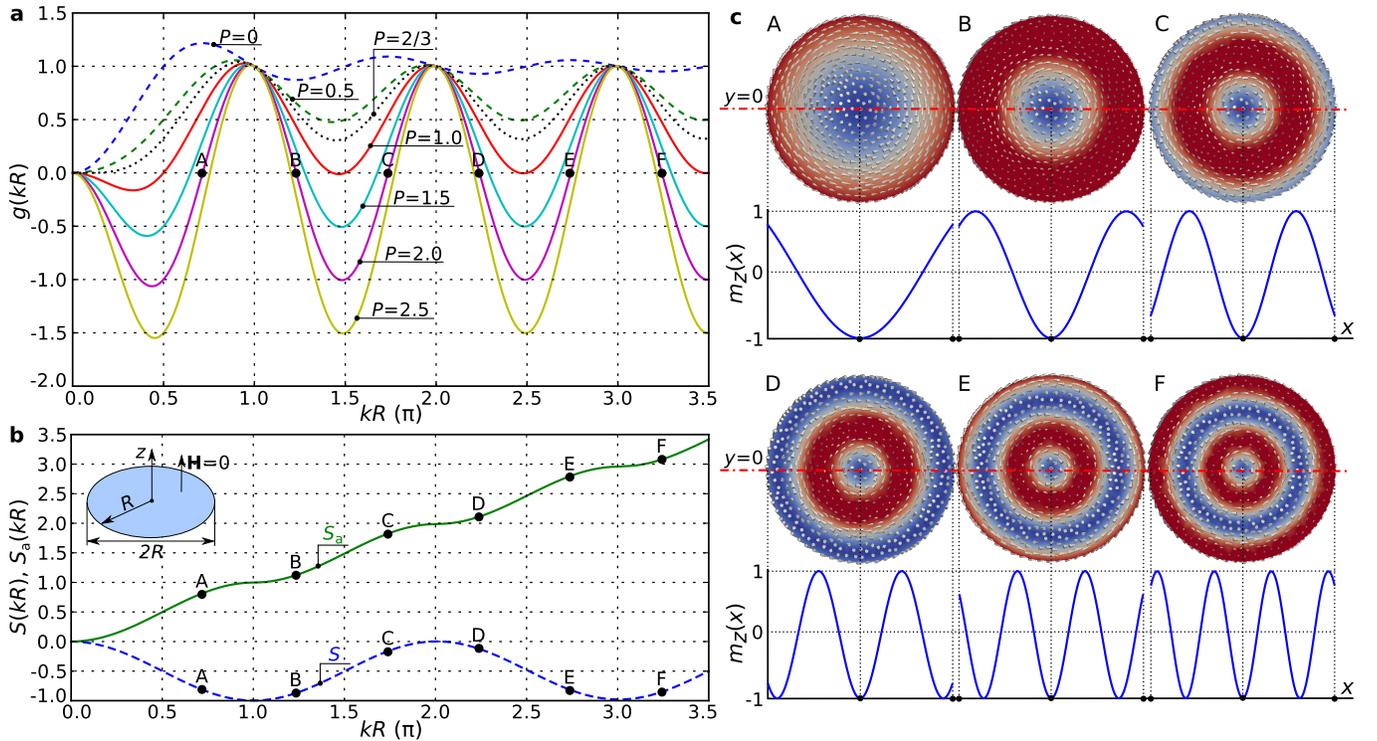

FIG. 2. **Zero-torque condition plots and magnetisation configurations corresponding to its solutions.** (**a**) The zero-torque condition $g(kR)$ plots, given by Eq. (5), for different values of $P = D/kA$. All zero-torque conditions for $P$ smaller than the boundary case $P = 2/3$ (dotted line) have no solutions, whereas for $P > 2/3$, multiple solutions (A-F) exist. (**b**) The non-injective dependence of skyrmion number $S$ and the injective dependence of scalar value $S_a$ on $kR$. (**c**) The magnetisation configurations and the out-of-plane magnetisation component $m_z(x)$ along the horizontal symmetry line for different solutions (A-F) of Eq. (5) for $P = 2$.

In addition to the previously defined eight chiral initial states, we also use the uniform magnetisation configuration, where the magnetisation at all mesh nodes is in the positive out-of-plane $z$-direction, as shown in Supplementary Fig. 1 marked as U. Finally, in order to capture other equilibrium states that cannot be obtained by relaxing previously described well-defined magnetisation configurations, we also use additional three random magnetisation configurations. At every mesh node, we choose three random numbers in the $[-1, 1]$ range for three magnetisation components and then normalise them in order to fulfil the $|\mathbf{m}| = 1$ condition. An example of one random magnetisation configuration is shown in Supplementary Fig. 1 and labelled as R.

## SUPPLEMENTARY SECTION S2: RELAXATION DIAGRAMS

In this section, we compute the equilibrium states at all points in the $d$–$H$ parameter space obtained by relaxing well-defined initial states, introduced in the Supplementary Section S1. More precisely, we compute the equilibrium states (local energy minima) that result from a particular initial condition. This allows us to provide a systematic overview of equilibrium states, and gain additional insight about the phase space energy landscape in the studied system. We vary the sample diameter between 40 nm and 180 nm in steps of $\Delta d = 4$ nm and the external magnetic field between $\mu_0 H = 0$ T and $\mu_0 H = 1.2$ T in steps of $\mu_0 \Delta H = 20$ mT. Relaxation diagrams are represented by plotting the scalar value $S_a$ as a function of disk diameter $d$ and applied field strength $H$, and Supplementary Fig. 3 shows the relaxation diagrams for skyrmionic initial configurations A-E, and Supplementary Fig. 4 shows the relaxation diagrams for helical H2, H3, H4, and uniform U initial configurations. The phase diagram of equilibrium states, shown in Fig. 1 in the main text, summarises the main results from the relaxation diagrams presented and discussed here.

We now discuss each of relaxation diagrams in Supplementary Fig. 3 and 4, where each subplot corresponds to one particular initial configuration. The scalar value $S_a$, as a function of disk sample diameter $d$ and external magnetic field

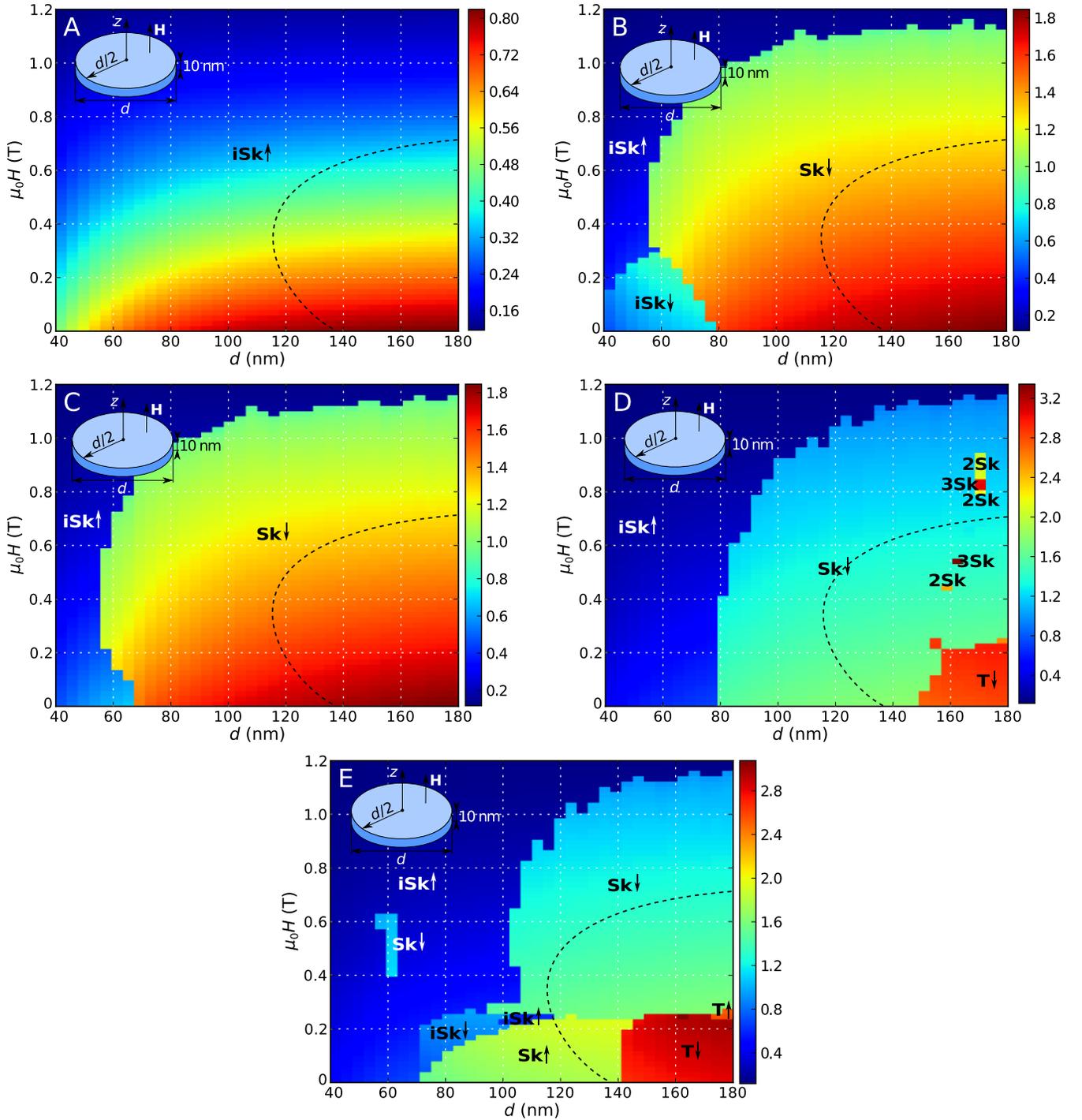

FIG. 3. **The relaxation diagrams obtained by relaxing skyrmionic initial state A-E.** The initial states correspond to the first five solutions of the analytic model and the phase diagrams are marked A-E accordingly. The relaxation diagrams are represented as the dependence of scalar value $S_a$ on the disk sample diameter $d$ and an external field $H$ (as shown in insets).

$H$, computed for the final relaxed equilibrium state, i.e. local or global energy minimum, we show in Supplementary Fig. 3 (A) for the energy minimisation process that started from the skyrmionic initial configuration A (shown in Supplementary Fig. 1). The "iSk↑" label refers to the incomplete Skyrmion (iSk) magnetisation configuration with the core magnetisation pointing in the positive $z$-direction. An example of this state, marked with the same label, is shown in Supplementary Fig. 5. From the Supplementary Fig. 3 (A), we can see that for all examined diameters and external field values, the final relaxed configuration is the iSk↑ state. Now, this relaxation diagram is compared with

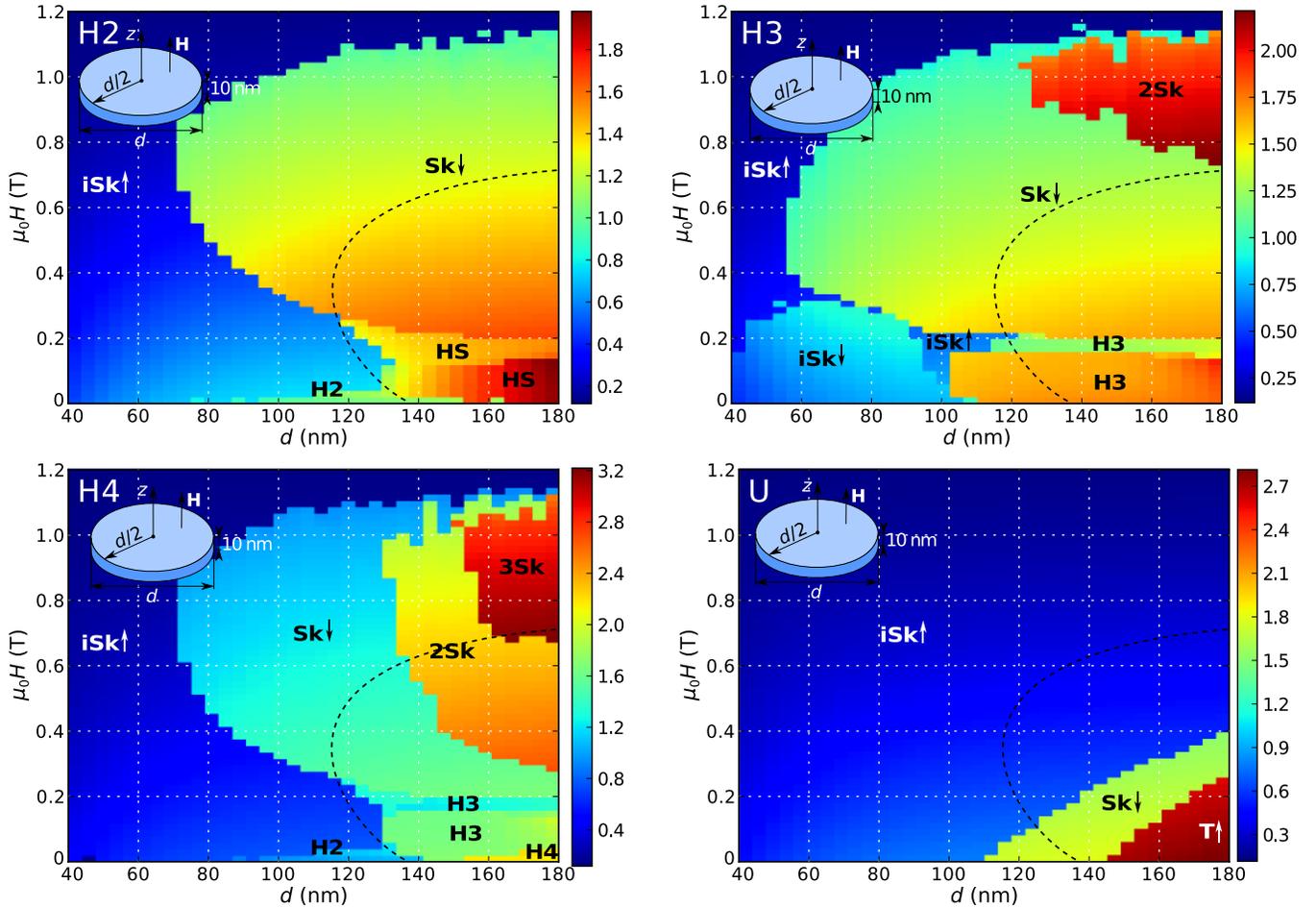

FIG. 4. **The relaxation diagrams obtained by relaxing helical and uniform initial state.** The initial states correspond to the helical states H2, H3, and H4, as well as the uniform state U, and the relaxation diagrams are marked accordingly. The relaxation diagrams are represented as the dependence of scalar value $S_a$ on the disk sample diameter $d$ and an external field $H$ (as shown in insets).

the ground state phase diagram (see Fig. 2 (a) in the main text) and the boundary between two ground state (iSk and Sk) regions is shown as the dashed line in the discussed relaxation diagram. We can see that the iSk is indeed the ground state (i.e. the global energy minimum) for $d < 140$ nm, but for larger diameters the isolated Skyrmion (Sk) configuration has a lower energy, and thus the iSk configuration is only a local energy minimum.

Similarly, if the energy minimisation process is started from the initial configuration B, shown in Supplementary Fig. 1 (B), the $S_a(d, H)$ for the final relaxed magnetisation state is obtained and shown in Supplementary Fig. 3 (B). The vast majority of the final configurations ($d \geq 80$ nm and $\mu_0 H \lesssim 1.1$ T), labelled as "Sk↓", correspond to the isolated skyrmion state with the core pointing in the negative $z$-direction. An example of this state we show in Supplementary Fig. 5, marked with the same label. By comparison with the ground state phase diagram shown in Fig. 2 (a) in the main text (dashed line), we can see that the Sk is the ground state for large diameters $d \geq 140$ nm and field values $\mu_0 H < 0.7$ T. However, for smaller diameters the isolated skyrmion configuration is only metastable (as the iSk configuration is the ground state). We can see that in the vicinity of $d \approx 60$ nm and $\mu_0 H \approx 0.1$ T parameter space point, the initial configuration B relaxes to the iSk configuration with core oriented in the negative $z$-direction (iSk↓), but for larger field values, configuration B falls into the iSk↑ configuration (see Supplementary Fig. 5 for detailed configurations of iSk↓ and iSk↑ states). The iSk↓ has a higher energy than the iSk↑ as the majority of the magnetisation is pointing in the direction opposite to the applied field. However, the initial configuration B is such that the core is pointing down, and there appears to be a direct energy minimisation path to the iSk↓ configuration for field values smaller than approximately 0.3 T. For larger fields, the Zeeman energy becomes so important that the initial configuration B leads to the the iSk↑ configuration.

Supplementary Fig. 3 (C) shows that the initial configuration C suppresses the iSk↓ state completely but is otherwise

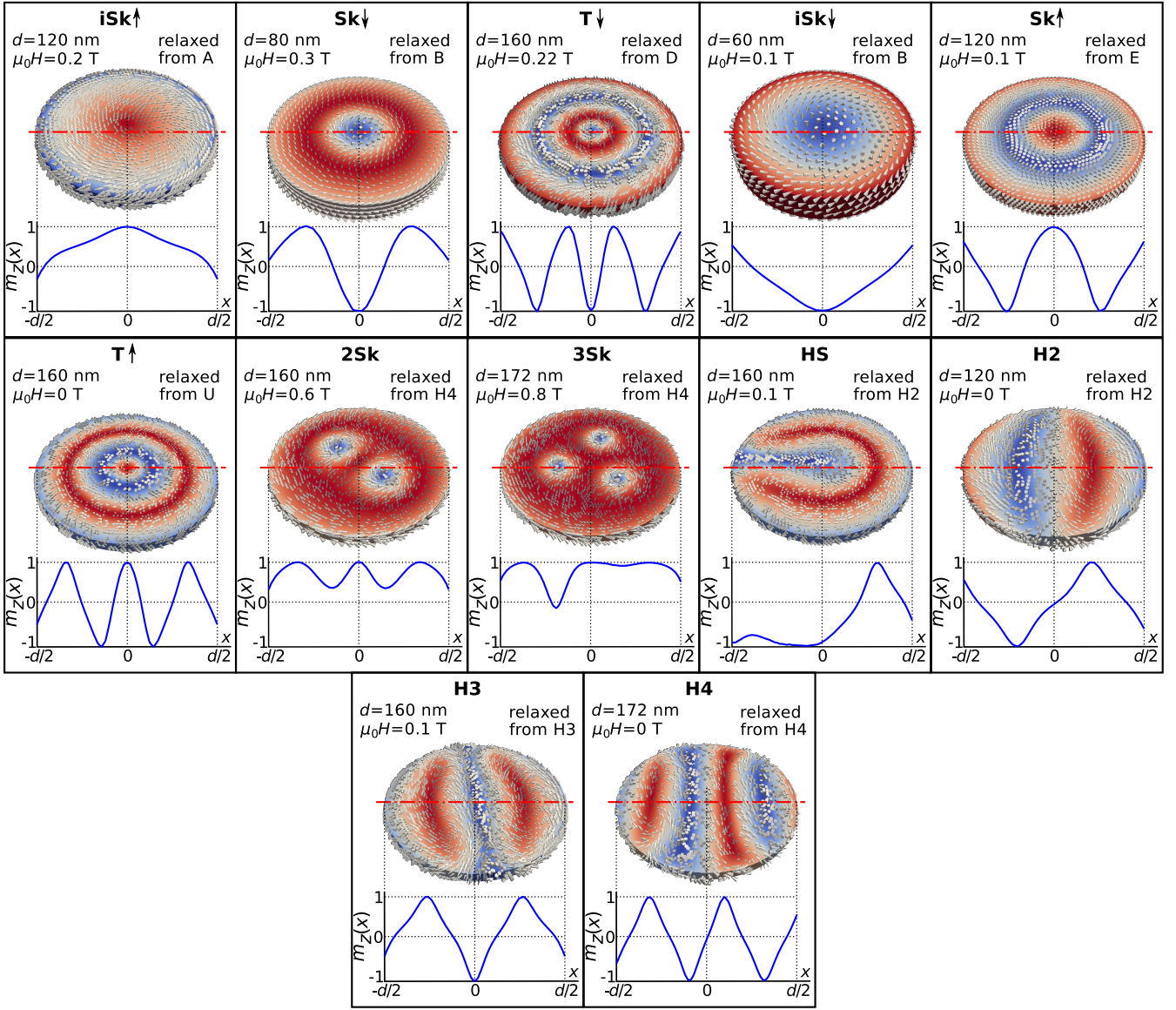

FIG. 5. **The identified equilibrium magnetisation configurations.**

similar to Supplementary Fig. 3 (B). We note in particular that the Sk configuration cannot exist for $d < 60$ nm even if the relaxation is started from a Sk-like configuration B or C, i.e. there are no metastable isolated skyrmion states at the smallest diameters.

If the system is relaxed from the initial state D, shown in Supplementary Fig. 1 (D), a qualitative change from Supplementary Fig. 3 (B) and (C) is evident as shown in Supplementary Fig. 3 (D). In addition to the iSk↑ and Sk↓ states, there are now a number of, according to Fig. 3 in the main text, higher energy metastable states emerging as 2 or 3 skyrmions in the disk (see Supplementary Fig. 5 for detailed plots of 2Sk and 3Sk states). Furthermore, for small field values $\mu_0 H \lesssim 0.2$ T and large diameters $d \gtrsim 152$ nm, the Target (T) equilibrium state with core orientation in the negative $z$-direction (T↓) arises. The T↓ state is shown in Supplementary Fig. 5.

The scalar value $S_a(d, H)$ computed for final equilibrium configurations in Supplementary Fig. 3 (E) we obtained by relaxing the initial configuration E shown in Supplementary Fig. 1. The initial skyrmionic state E does not relax to 2 and 3 skyrmion configurations but allows the Sk↑ state to arise for small field values.

Supplementary Fig. 4 (H2), (H3) and (H4), show the $S_a(d, H)$ for starting configurations H2, H3 and H4, respectively, as shown in Supplementary Fig. 1. All three initial configurations result in the incomplete Skyrmion configuration with core pointing up (iSk↑) for the smallest diameters as well as for largest fields. The H3 initial

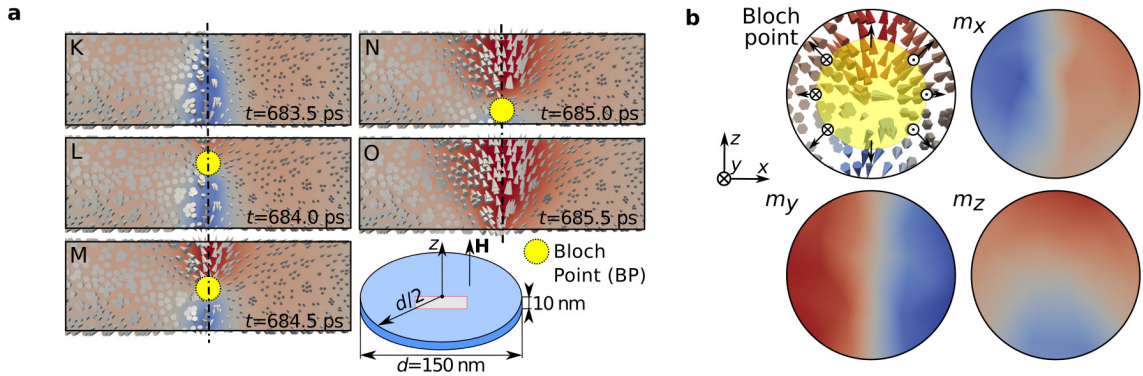

FIG. 6. **The isolated skyrmion orientation reversal in confined three-dimensional helimagnetic nanostructure with downwards Bloch point propagation direction.** (**a**) The $m_z$ colourmap and magnetisation field in the central part of $xz$ cross section as shown in an inset together with the position of Bloch Point (BP). (**b**) The BP structure along with colourmaps of magnetisation components which shows that the magnetisation covers the closed surface (sphere surrounding the BP) exactly once.

configuration relaxes into a configuration with two Skyrmions in the disk (2Sk) for $d > 120\,\text{nm}$ and field values between $0.8\,\text{T}$ and $1.1\,\text{T}$. These 2Sk configurations had appeared occasionally when starting from configuration D (see Supplementary Fig. 3 (D)). The H4 initial configuration also encourages 3 skyrmions in the disk to arise as a metastable state.

Supplementary Fig. 5 (U) shows $S_a$ for final configurations when the simulation starts from a uniform magnetisation, pointing up in the positive $z$-direction. This results mostly in the incomplete Skyrmion configuration with core pointing up (iSk↑). However, we also find the Skyrmion with core pointing down Sk↓ and the Target configuration T↑ as the diameter increases and the field decreases. Supplementary Fig. 3 (A) is interesting to compare with Supplementary Fig. 4 (U): in the former, only the iSk↑ state results, presumably because from the initial state A, only the iSk↑ state is accessible in the relaxation. On the contrary, for the uniform configuration, the system finds the energy minimum for the isolated Skyrmion state Sk↓ and the Target T↑ because other energy minima can be accessed from this initial state. Fig. 5 (a) in the main text shows the relative energies of the different metastable states for $H = 0$.

## SUPPLEMENTARY SECTION S3: DIFFERENT BLOCH POINT PROPAGATION DIRECTION

In order to illustrate a different direction of the Bloch Point (BP) propagation, we show a result from another skyrmion reversal. The simulation parameters are the same as in Fig. 7 of the main text, except that the Gilbert damping $\alpha$ is increased from 0.3 to 0.35. We show the results of isolated skyrmion core orientation reversal dynamics with modified Gilbert damping in Supplementary Fig. 6.

Now, the obtained reversal dynamics is compared with the reversal dynamics shown in Fig. 7 in the main text. From the Supplementary Fig. 6 (a), we can see that the Bloch point enters the sample at the top boundary at approximately 684 ps, then propagates downwards to the bottom boundary, where it leaves the sample at approximately 685 ps. Because of the opposite BP propagation direction, the structure of the Bloch Point changes (compare Fig. 7 (e) in the main text with Supplementary Fig. 6 (b)). More precisely, the out-of-plane magnetisation component $m_z$ field in the vicinity of BP is changed so that for the upper half of BP $m_z > 0$, whereas in the lower half $m_z < 0$.

## SUPPLEMENTARY SECTION S4: HIGHER ORDERING TEMPERATURE MATERIAL

The ordering temperature of simulated FeGe material, $T_C = 278.7\,\text{K}$ [4], is lower than the room temperature, which means that this material cannot be used to fabricate a device operating at room temperature. Therefore, it is important to determine how our results regarding the identified lowest energy state would change for the material with higher ordering temperature. Because no high ordering temperature helical B20 material has been reported to this day, the best we can do is to artificially increase the ordering temperature, estimate new material parameters, and repeat the study of equilibrium states.

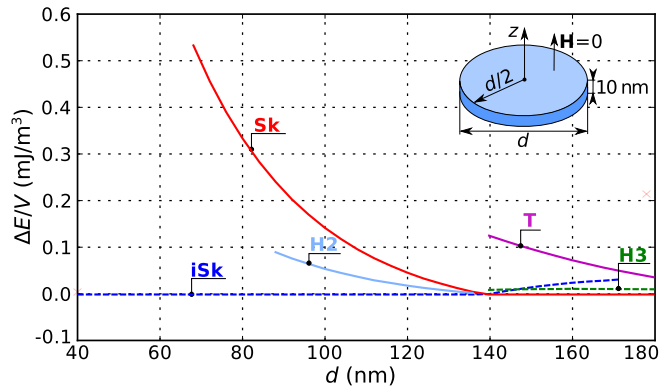

FIG. 7. **The energy density differences between all identified equilibrium states and corresponding lowest energy state as a function of disk sample diameter.** A full set of initial state configurations are relaxed for different disk sample diameter values at zero external magnetic field. The helimagnetic material ordering temperature is artificially increased to $T_C = 350\,\mathrm{K}$ (above room temperature).

We increase the ordering temperature to $T_C = 350\,\mathrm{K}$, and calculate new values of exchange and Dzyaloshinskii-Moriya energy constants (following the estimation described in Methods section of the main text) and obtain $A = 1.1 \times 10^{-11}\,\mathrm{J/m}$ and $D = 1.98 \times 10^{-3}\,\mathrm{J/m^2}$. Using these values, we relax the full set of initial magnetisation configurations at zero external magnetic field for different disk sample diameters. More precisely, we vary the disk sample diameter between $40\,\mathrm{nm}$ and $180\,\mathrm{nm}$ in steps of $4\,\mathrm{nm}$ and compute the energy density $E/V$ of all identified relaxed equilibrium states, where $V$ is the sample volume. After that, from the computed energy density, we subtract the energy density of the corresponding lowest energy state. We plot the calculated energy density differences of all equilibrium states as a function of disk sample diameter $d$ and show them in Supplementary Fig. 7.

By comparing the Supplementary Fig. 7 for higher ordering temperature material with Fig. 3 in the main text, we conclude that the incomplete Skyrmion (iSk) state remains being the lowest energy state for $d < 140\,\mathrm{nm}$, whereas the isolated Skyrmion (Sk) state is the lowest energy state for $d \geq 140\,\mathrm{nm}$. However, the iSk state is not in equilibrium for disk sample diameters larger than $172\,\mathrm{nm}$, which is in contrast to the FeGe material case where iSk is in equilibrium for all examined $d$ values. Another difference is that the metastable Target (T) state is the highest energy state for the whole $d$ range where it is in equilibrium. Finally, we do not observe H4 state (helical state with four half periods) for this high ordering temperature material.

## SUPPLEMENTARY REFERENCES

---

* mb4e10@soton.ac.uk
† h.fangohr@soton.ac.uk


\end{document}